\documentclass[10pt,letterpaper]{article}
\usepackage[top=0.85in,left=1.1in,footskip=0.75in]{geometry}

\usepackage{floatrow}           
\usepackage{graphicx}
\usepackage{dcolumn}
\usepackage{array}
\usepackage[english]{babel}
\newcolumntype{P}[1]{>{\centering\arraybackslash}p{#1}} 
\newcolumntype{R}[1]{>{\raggedright\arraybackslash}p{#1}} 

\usepackage{amsmath,amssymb}
\usepackage{changepage}
\usepackage[utf8x]{inputenc}
\usepackage{bbm}
\usepackage{cite}
\usepackage[right]{lineno} 
\usepackage{microtype}
\DisableLigatures[f]{encoding = *, family = * }

\newlength\savedwidth

\usepackage{setspace}

\usepackage[aboveskip=1pt,labelfont=bf,labelsep=period,justification=raggedright,singlelinecheck=off]{caption}

\makeatletter
\renewcommand{\@biblabel}[1]{\quad#1.}
\makeatother

\date{}

\usepackage{lastpage,fancyhdr}
\usepackage[]{graphicx} 
\usepackage[outdir=./]{epstopdf}
\usepackage{bm}
\usepackage{color}
\usepackage{bbold}
\usepackage{soul}
\usepackage{epsfig,epic}
\usepackage{placeins}

\newcommand{\FIG}[1]{Fig~\ref{fig:#1}}
\newcommand{\FIGSUPP}[1]{Fig~\ref{fig:#1}}

\newcommand{\SIref}[1]{Sec.~\ref{#1}}
\newcommand{\SI}[1]{Sec.~\ref{#1}}

\newcommand{\newc}{\newcommand}
\newc{\beq}{\begin{linenomath}\begin{equation}}
\newc{\eeq}{\end{equation}\end{linenomath}}
\newc{\kt}{\rangle}
\newc{\br}{\langle}
\newc{\beqa}{\begin{eqnarray}}
\newc{\eeqa}{\end{eqnarray}}
\newc{\longra}{\longrightarrow}
\renewcommand{\eqref}[1]{Eq.~(\ref{eq:#1})}
\usepackage{url}

\newcommand{\leri}[1]{\left( #1 \right)}
\newcommand{\lerii}[1]{\left[ #1 \right]}
\newcommand{\leriii}[1]{\left| #1 \right|}
\newcommand{\leriv}[1]{\left< #1 \right>}

\newcommand{\abs}[1]{\leriii{#1}}
\newcommand{\norm}[1]{\abs{#1}}

\newcommand{\erfc}{\mathrm{erfc}}
\newcommand{\arccot}{\mathrm{arccot}}
\newcommand{\sgn}{\mathrm{sgn}}
\newcommand{\Int}{\int\limits}

\newcommand{\id}[1]{\mathrm d #1 \,}
\newcommand{\dx}[2]{\partial_{#2} #1} 
\newcommand{\dt}[1]{\dot{#1}}
\newcommand{\ddt}[1]{\ddot{#1}}
\renewcommand{\vec}[1]{\mathbf{#1}}
\newcommand{\mat}[1]{\mathbf{#1}}
\newcommand{\Nabla}{\bm{\nabla}}
\renewcommand{\phi}{\varphi}
\newcommand{\sign}[1]{\ensuremath{\text{sgn}\leri{#1}}}
\newcommand{\im}{\mathrm i}
\newcommand{\x}{X}
\newcommand{\y}{Y}

\renewcommand{\t}{T}
\newcommand{\utau}{\tau}
\newcommand{\xb}{\hat{x}}
\newcommand{\yb}{\hat{y}}
\newcommand{\zb}{\hat{z}}
\newcommand{\rb}{\hat{r}}
\newcommand{\vb}{\hat{v}}
\newcommand{\ub}{\vec{\hat{u}}}
\newcommand{\gam}{\gamma}  
\newcommand{\geofactor}{g} 
\newcommand{\curv}{\kappa} 
\newcommand{\tor}{\tau} 
\newcommand{\re}{\ensuremath{\text{Re}}} 

\newcommand{\nkol}{\ensuremath{\eta_{\text{Kol}}}} 
\newcommand{\tkol}{\ensuremath{\tau_{\text{Kol}}}} 
\newcommand{\rate}{\ensuremath{q}} 
\newcommand{\shearrate}{shear rate}
\newcommand{\shear}{\ensuremath{\alpha}} 
\newcommand{\shearopt}{\ensuremath{\shear^*}}
\newcommand{\D}{\ensuremath{D}} 
\newcommand{\success}{fertilization probability}
\newcommand{\successs}{encounter probability}
\newcommand{\prob}{\ensuremath{P}} 
\newcommand{\suc}{\ensuremath{\prob_{\text{fert}}}} 
\newcommand{\su}{\ensuremath{\prob_{\text{sperm:egg}}}} 
\newcommand{\sucfert}{\ensuremath{p_{\text{f}}}} 

\newcommand{\eps}{\ensuremath{\epsilon}}
\newcommand{\epsopt}{\ensuremath{\eps^*}}
\newcommand{\cb}{c_b} 
\newcommand{\Rmax}{\ensuremath{r_{\text{max}}}} 
\renewcommand{\k}{\ensuremath{f}}
\newcommand{\tmax}{\ensuremath{t_{\text{max}}}}

\newcommand{\Regg}{r_{\text{egg}}} 
\newcommand{\h}{\vec{h}} 
\newcommand{\vf}{\ensuremath{\vec{v}_{\text{ext}}}} 
\newcommand{\vfb}{\ensuremath{\vec{\vb}_{\text{ext}}}} 
\newcommand{\vo}{\ensuremath{v_0}}
\newcommand{\vs}{\ensuremath{v_{h}}} 
\newcommand{\bet}{v_{\phi}}   
\newcommand{\G}{\ensuremath{G}} 
\newcommand{\e}{\vec{e}}               
\newcommand{\psperm}{\ensuremath{\rho}} 
\newcommand{\lam}{k}              
\newcommand{\sig}{\sigma}
\newcommand{\relrate}{\ensuremath{\dt{Q}}} 
\newcommand{\ratio}{\ensuremath{a_{y}}} 
\newcommand{\ptO}{\ensuremath{p_{0,t_0}}}
\newcommand{\ptI}{\ensuremath{p_{1,t_0}}}
\newcommand{\ptII}{\ensuremath{p_{2,t_0}}}
\newcommand{\pyO}{\ensuremath{p_{0,y_0}}}
\newcommand{\pyI}{\ensuremath{p_{1,y_0}}}
\newcommand{\ymax}{\ensuremath{\y(c = \cb)}}

\begin{document}
\vspace*{0.2in}

\begin{flushleft}
{\Large
\textbf{Sperm chemotaxis in marine species is optimal at physiological flow rates according theory of filament surfing
}\\[1mm]
}
Steffen Lange\textsuperscript{1,2*},
Benjamin M. Friedrich\textsuperscript{2,3}
\\
\bigskip
\textbf{1} HTW Dresden, Dresden, Germany
\\
\textbf{2} Center for Advancing Electronics Dresden, TU Dresden, Germany
\\
\textbf{3} Cluster of Excellence Physics of Life, TU Dresden, Germany
\bigskip

* steffen.lange@tu-dresden.de\\
\end{flushleft}

\section*{Abstract} 
Sperm of marine invertebrates have to find eggs cells in the ocean. Turbulent flows mix sperm and egg cells up to the millimeter scale; below this, active swimming and chemotaxis become important. Previous work addressed either turbulent mixing or chemotaxis in still water. Here, we present a general theory of sperm chemotaxis inside the smallest eddies of turbulent flow, where signaling molecules released by egg cells are spread into thin concentration filaments. Sperm cells `surf' along these filaments towards the egg. External flows make filaments longer, but also thinner. These opposing effects set an optimal flow strength. The optimum predicted by our theory matches flow measurements in shallow coastal waters. Our theory quantitatively agrees with two previous fertilization experiments in Taylor-Couette chambers and provides a mechanistic understanding of these early experiments. `Surfing along concentration filaments' could be a paradigm for navigation in complex environments in the presence of turbulent flow.




\section*{Introduction}

Chemotaxis - the navigation of biological cells guided by chemical
gradients - is crucial for bacterial foraging, neuronal development,
immune responses, and sperm-egg encounter during
fertilization~\cite{FraGun1961,BerBro1972,DevZig1988,AlvFriGomKau2014,EisGio2006}.
Despite a century of research, most studies assumed perfect
concentration gradients of signaling molecules. Yet, in natural
environments, concentration fields of these chemoattractants are
non-ideal, distorted e.g. by turbulent flows. An unusually accessible
model system of such a cellular navigation is the chemotaxis of
sperm cells in marine invertebrates with external fertilization. For fertilization, sperm cells of many
species are known to employ chemotaxis to steer up concentration
gradients of signaling molecules released by the egg.
This sperm chemotaxis has been
intensively studied for external fertilization of marine
invertebrates, where sperm and egg cells are spawned directly into the
sea~\cite{Mil1985,EisGio2006,RifKruZim2004,CorTabWooGueDar2008,Kau2012}.
In this case, sperm and egg cells become strongly diluted. Besides
synchronized spawning~\cite{SerPeaKauBra1996,GorBra2004}, sperm
chemotaxis is important to enhance sperm-egg encounter
rates~\cite{Lev1993}. The mechanism of sperm chemotaxis in marine
invertebrates is well established
theoretically~\cite{Cre1996,FriJue2007} and has been experimentally
confirmed~\cite{JikAlvFriWilPasColPicRenBreKau2015}: Sperm cells swim
along helical paths $\vec{r}(t)$, while probing the surrounding
concentration field $c(\vec{r})$. A cellular signaling system rotates
the helix axis $\h$ to align with the gradient $\Nabla c$ at a rate
proportional to a normalized gradient gradient strength
$\norm{\Nabla c}/\leri{c+\cb}$ reflecting sensory adaption with
sensitivity threshold
$\cb$~\cite{KasAlvSeiGreJaeBeyKraKau2012,KroMaeLanBaiFri2018}.

\begin{figure}[t]
  \includegraphics{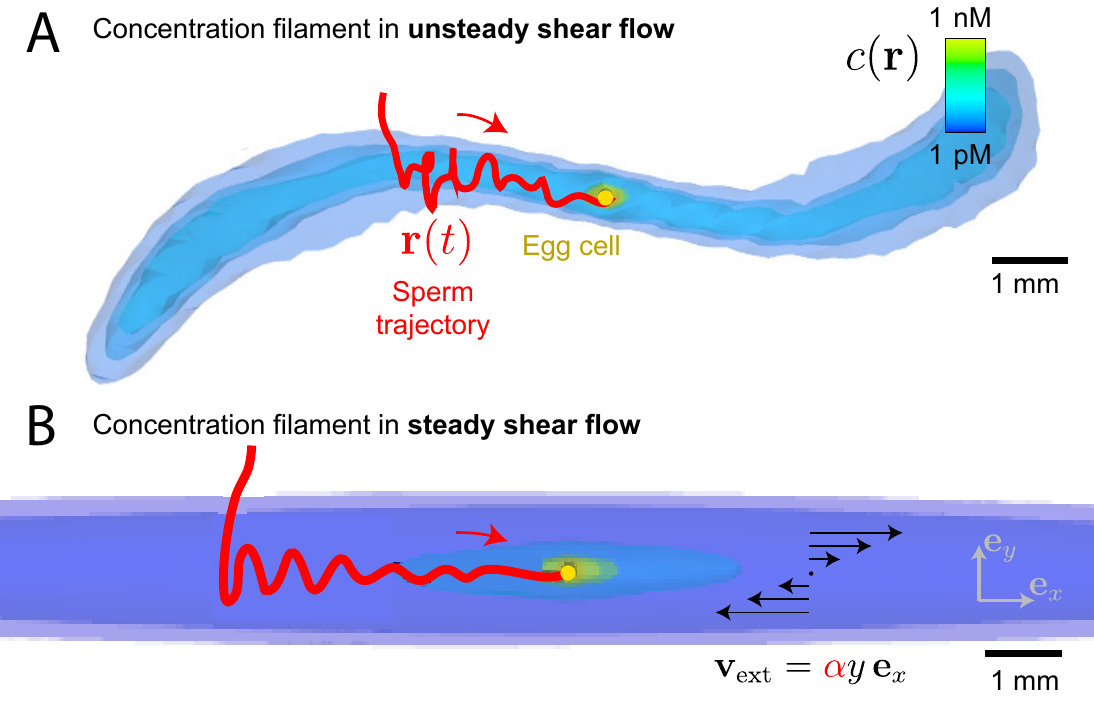}
  \caption{\label{fig:filament-surfing} \textbf{Sperm chemotaxis in
      external flow.} (A) Simulated, three-dimensional concentration
    field $c(\vec{r})$ of chemoattractant released from a
    freely-rotating, spherical egg (yellow sphere) suspended in
    unsteady shear flow as a model of small-scale turbulence. An
    exemplary simulated sperm cell (trajectory in red) finds the
    elongated concentration filament by chance and subsequently
    `surfs' along the filament by chemotaxis (B) Same as (A), but for
    the prototypical idealization of simple shear flow
    $\vf(\vec{r}) = \shear y \, \e_x$ accounting for convection and
    co-rotation by the external flow. We obtain a generic form of the
    concentration filament, \eqref{filament-main}, and characterize
    surfing along the filament analytically as a damped oscillation.
    Parameters correspond to sea urchin \emph{A. punctuala}, assuming
    continuous release of chemoattractant at constant rate
    $\relrate = 0.46~\text{fmol min}^{-1}$ for an exposure time
    $\tmax = 6~\text{min}$. Constant \shearrate{}
    $\shear = 0.17~\text{s}^{-1}$ in (B), corresponding to
    root-mean-square \shearrate{} of (A). Same color-code for
    concentration in (A) and (B), but different level sets. We use a
    generic theoretical description of helical sperm chemotaxis, see
    \emph{Methods and Materials} for details (helix radius
    $r_0 = 7~\mu\text{m}$ not visible at length-scale of figure).}
\end{figure}

Previous work on sperm chemotaxis focused predominantly on idealized
conditions of still water~\cite{Eis1999,Kau2012}. However, natural
habitats like the ocean are characterized by turbulent flow, which
convects and co-rotates gametes and distorts concentration fields into
filamentous
plumes~\cite{RifKruZim2004,RifZim2007,ZimRif2011,TaySto2012,KasAlvSeiGreJaeBeyKraKau2012,CriZim2014,RusGuaSto2014},
see \FIG{filament-surfing}A for illustration. Turbulence in
typical spawning habitats of marine invertebrates has been
characterized, e.g., in terms of local energy dissipation rates per
mass
$\eps = 10^{-9} -
10^{-6}~\text{m}^2\text{s}^{-3}$~\cite{LazMan1989,MeaDen1995,RifZim2007,JumTroBosKar2009,CriZim2014}
corresponding to typical \shearrate{}s
$\alpha=0.03 - 1~\text{s}^{-1}$, which are often similar to those in
mammalian reproductive tracts~\cite{EisGio2006}. Turbulent flow
rapidly mixes sperm and egg cells, yet only down to the Kolmogorov
length-scale $\nkol = \leri{\nu^3/\eps}^{1/4} = 1-10~\text{mm}$ (with
kinematic viscosity $\nu$). Previous predictions based on turbulent
mixing~\cite{Sre2019} substantially underestimated \success{}
$\suc$~\cite{DenShi1989,CriZim2014}, since these early studies
neglected active swimming and sperm chemotaxis inside the smallest
eddies, whose size is comparable to the Kolmogorov length $\nkol$. At
these small length-scales, the Reynolds number of the flow is below
one, and gametes perceive turbulence as unsteady shear
flow~\cite{LazMan1989,JumTroBosKar2009} with a typical \shearrate{}
$\shear$ set by the inverse of the Kolmogorov time
$\tkol = \sqrt{\nu/\eps}$. Intriguingly, fertilization experiments
conducted at physiological \shearrate{}s hint at the existence of an
optimal \shearrate{} $\shearopt>0$, corresponding to an optimal
turbulence strength $\epsopt>0$, at which the \success{} \suc{} was
maximal~\cite{MeaDen1995,ZimRif2011,BelCri2015}. Similar observations
have been made in direct numerical simulations of bacterial
chemotaxis~\cite{TaySto2012}. Obvious biological effects can be ruled
out as the origin of the optimum~\cite{MeaDen1995,RifZim2007},
including flow damaging the gametes or sperm-egg bonds being broken by
shear forces. Despite an early two-dimensional
model~\cite{BelCri2015}, a physical explanation and quantitative
understanding of the observed optimum is still
missing~\cite{ZimRif2011, CriZim2014}.

Here, we develop a theory of sperm chemotaxis in small-scale
turbulence: As a prototypical model, we consider sperm chemotaxis in
simple, three-dimensional shear flow, which convects and co-rotates
sperm cells and distorts the chemoattractant field that surrounds the
egg. We predict an optimal \shearrate{} \shearopt{} in simulations, as
previously suggested by experiments~\cite{MeaDen1995,ZimRif2011}. We
provide a novel mechanistic explanation of this optimum from theory:
We describe how external flow distorts concentration fields into
slender filaments, and how sperm cells `surf' along these filaments
towards the concentration source, see \FIG{filament-surfing}B. The
optimum arises from the competition between accelerated spreading of
the chemoattractant at increased flow, which enhances chemotaxis, and
filaments becoming increasingly thinner, which impairs chemotaxis. We
apply our theoretical description to two previous experiments on sperm
chemotaxis, one with moderate flow, mimicking fertilization in shallow
coastal waters~\cite{ZimRif2011}, and one with strong turbulence,
mimicking fertilization in the surf zone~\cite{MeaDen1995}. In both
cases, simulation and theory match the experimental data, yet also
prompt a partial re-interpretation of these early experiments: We
infer a high background concentration of chemoattractant in these
experiments, which actually masks the existence of an optimal flow
strength for the experimental conditions used (in contrast to physiological
spawning habitats where no relevant background concentration should be
present). We propose that `surfing along concentration filaments'
could be a common navigation paradigm in natural habitats
characterized by external flows, which is relevant for the last
millimeters towards a source.

\section*{Results}

\subsection*{Simulations: Optimal \shearrate{}}

\begin{figure}[t]
\centering
\includegraphics{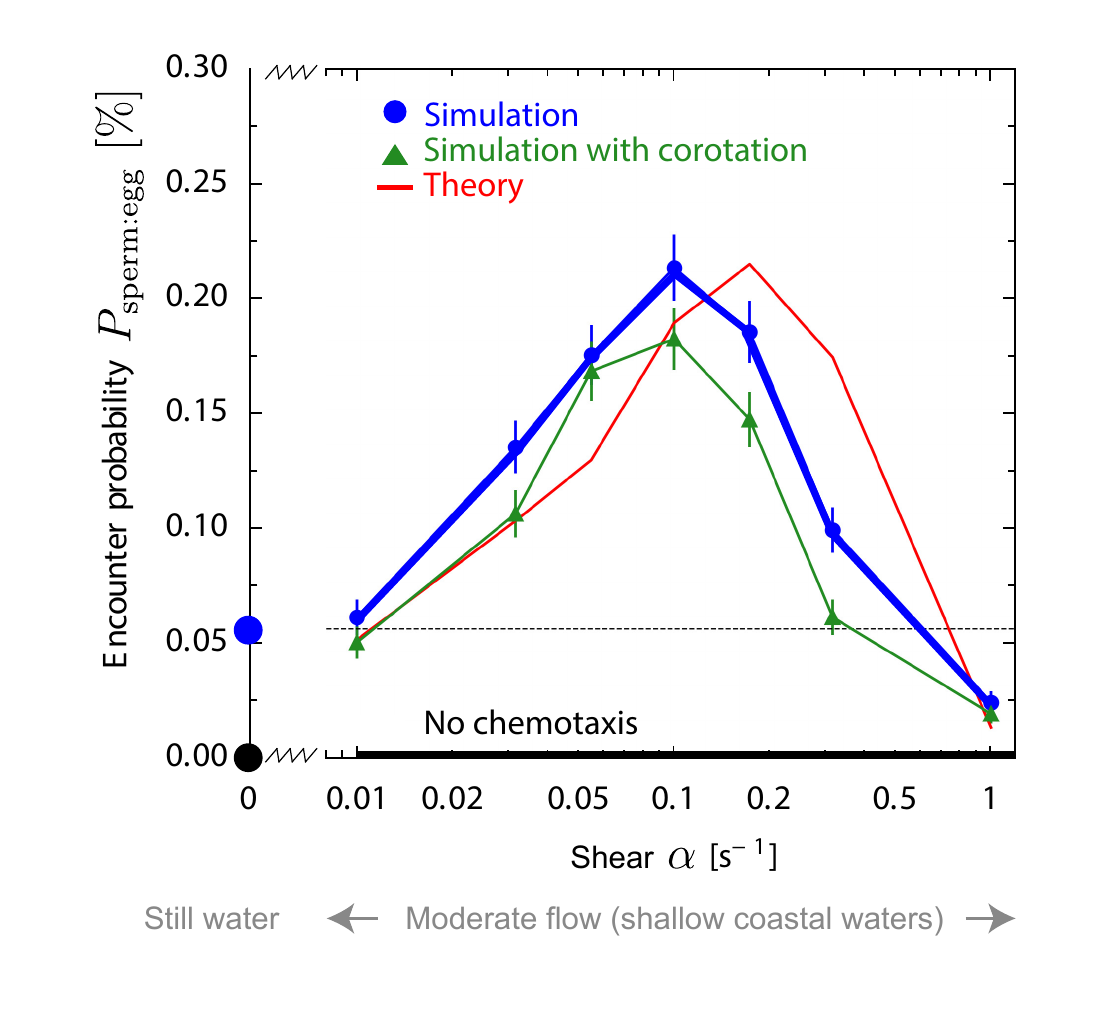}
\caption{\label{fig:suc} \textbf{Sperm-egg-\successs{} displays maximum
    as function of \shearrate{} in simulations for sea urchin sperm at
    physiological flow rates.} Probability $\su(\shear)$ that a single
  sperm cell finds an egg as function of external \shearrate{}
  $\shear$. Simulations account for flow-induced distortion of
  concentration fields into long filaments as well as convection and
  co-rotation of sperm cells by the flow (green triangles, mean $\pm$
  SD). Without co-rotation results change only marginally (blue
  circles). Simulation results agree with predictions from our theory
  of \emph{filament surfing} (red, presented below). Without sperm
  chemotaxis, the encounter probability is virtually zero
  ($< 10^{-5}$, black). Our theory has a single fit parameter, the
  flux of sperm cells arriving at the filament,
  $j_{\text{out}} = 0.063~\text{m}^{-2}\text{s}^{-1}$. This value
  matches in magnitude the limit
  $j_{\text{out}} = \rho_{\text{egg}} \vs/4 =
  0.04~\text{m}^{-2}\text{s}^{-1}$ for a ballistic swimmer with random
  initial conditions, see \SIref{sec:effective-volume} for details.
  Parameters as in \FIG{filament-surfing}B.}
\end{figure}

We simulate sperm chemotaxis in a simple shear flow
$\vf\leri{\vec{r}} = \shear y \, \e_x$,
extending a generic theory of
helical chemotaxis~\cite{FriJue2007} by incorporating convection and
co-rotation of sperm cells by the external fluid flow. In particular,
for co-rotation by the flow the Jeffery
equation~\cite{Jef1922,PedKes1992} is employed.
We ask for encounters of sperm cells with a single egg that releases chemoattractant molecules,
which establish a concentration field $c(\vec{x},t)$ by convection and diffusion.
By switching to a co-moving frame in which the egg is at rest.
we may assume that the suspended egg is located at the origin
$\vec{r}=\vec{0}$
without loss of generality.
We use a spherical
periodic boundary at radius $\Rmax$, which mimics an ensemble of eggs
with density $\rho_{\text{egg}} = \leri{4 \pi \Rmax^3/3}^{-1}$, and
assume an exposure time $\tmax${}.
For turbulent flow, the exposure time would correspond to a typical time interval between subsequent intermittency events
that re-mix sperm and egg cells and reset any concentration field of
chemoattractant that might have been established in between.
(\textit{Methods and Materials} provides
details on simulation setup and extensive discussion of parameters.)
The resulting sperm-egg-\successs{} \su{} displays a maximum at an
optimal \shearrate{} $\shearopt\approx 0.1~\text{s}^{-1}$, see
\FIG{suc}, which uses parameters for sea urchin \emph{A.
  punctuala}. At the optimal \shearrate{} $\shearopt$, \su{} is
$4$-fold higher than without flow. Only for larger \shearrate{}s
$\shear>0.3~\text{s}^{-1}$, chemotaxis becomes less effective than
without flow and finally ineffective at very strong \shearrate{} with
$\shear\geq1~\text{s}^{-1}$. Note that without chemotaxis, the
\successs{} is 2-3 orders of magnitude smaller for the chosen
parameters (not shown as not visible).

Surprisingly, the numerical results show that co-rotation of sperm
cells is not necessary for the existence of an optimal \shearrate{} as
simulations without co-rotation yield very similar results, see
\FIG{suc}. Consequently, the existence of an optimal \shearrate{}
$\shearopt$ should be a consequence of the distortion of the
concentration field by the flow. For simplicity, we thus focus on the
case without co-rotation in the following (simulations with
co-rotation are displayed in Figs~\ref{fig:fig5c-coro} and
\ref{fig:fig5-coro} ). Typically, shear flow
generates long filaments, or plumes, of high concentration.
Simulations show how sperm cells enter these filaments and `surf'
along them, see \FIG{filament-surfing}B, with trajectories resembling
a damped oscillation, see also \FIGSUPP{damped-oscillation}. Damped
oscillations occur when sperm cells move towards the egg, yet
oscillations are amplified when sperm cells move away from the egg.
The latter sometimes causes sperm cells swimming in the wrong
direction to turn around, thus redirecting them towards the egg. In
conclusion, sperm chemotaxis in external flows is a two-stage search
problem~\cite{HeiMar2020} of first finding a concentration filament
and subsequent chemotactic surfing along this filament towards the
egg.

\subsection*{Theory: Filament surfing}

We develop a theory of sperm chemotaxis in filamentous concentration
fields generated by simple shear flows. This theory describes surfing
along filaments and allows to predict the sperm-egg-\successs{}, see
\FIG{suc}. We consider a simple shear flow
$\vf\leri{\vec{r}} = \shear y \, \e_x$ and a spherical egg of radius
$\Regg$, without loss of generality located at $\vec{r}=\vec{0}$,
releasing chemoattractant at a constant rate for a time $t$. The
choice of the coordinate system corresponds to a co-moving frame in
which the egg is at rest. Far from
the source $\norm{\vec{r}}\gg \Regg$, the concentration field
$c(\vec{r},t)$ established by diffusion and convection takes a generic
form, see \FIG{filament-surfing}B for illustration,
\begin{linenomath}\begin{align}
\label{eq:filament-main}
  c(\vec{r},t) = c_0 \exp\leri{-\lam \abs{x}}
  \exp\leri{-\frac{\leri{y-y_0}^2/\ratio^2 + z^2}{2 \sig^2}} \ .
\end{align}\end{linenomath}
\eqref{filament-main} describes a filament with exponential decay
along its center line $\leri{x,y_0(x,t),0}$ and a Gaussian
cross-sectional profile. We derived phenomenological power-laws for
all parameters $c_0(t)$, $\lam(t)$, $\sig(\abs{x},t)$, $\ratio$, and
$y_0(x,t)$, see \SIref{sec:filament-anal} for details. Importantly,
the effective decay length along the centerline of the concentration
filaments increases with flow, $1/\lam \sim \shear$, while the
effective diameter of the filaments decreases with flow, since the
decay length $\sigma$ away from the centerline of the filaments is
independent of the flow while the base concentration $c_0$ decreases
with flow, $c_0 \sim \shear^{-1}$. In this sense, the concentration
filaments become longer and thinner with increasing \shearrate{}
$\shear$.

Sperm cells from marine invertebrates swim along helical paths,
along which they measure the local concentration of chemoattractant \cite{JikAlvFriWilPasColPicRenBreKau2015}.
This time-dependent concentration signal exhibits characteristic oscillations at the frequency of helical swimming,
which encode direction and strength of a local concentration gradient.
The concentration signal elicits a continuous steering response by which the helical swimming path aligns with the gradient.
We generalize an effective equation for the alignment of the helix
axis $\h(t)$ with the local gradient $\Nabla c(\vec{r}(t))$,
previously derived for simple radial concentration
fields~\cite{FriJue2007},
\begin{linenomath}\begin{align}
  \dt{\Psi} &= - \bet \frac{\norm{\Nabla c}}{c + \cb} \sin \Psi \ , \qquad \Psi = \sphericalangle \leri{\Nabla c,\h} \ ,
\end{align}\end{linenomath}
with an effective response parameter $\bet$ of chemotactic signaling,
to concentration filaments given by \eqref{filament-main}, see
\SIref{sec:chemotactic-navigation} for details. For a
normalized distance $\y$ of the sperm trajectory from the centerline
of the concentration filament, we obtain a one-dimensional effective
equation of motion which explains and quantifies filament surfing,
\begin{linenomath}\begin{align}
  \label{eq:dimless-ODE-main}
  \ddt{\y} = \big(\underbrace{-\leri{1-\dt{\y}^2}\y}_{\sim\text{oscillator}} \pm
  \underbrace{\gam
  \sqrt{\leri{1-\dt{\y}^2}}\dt{\y}}_{\sim\text{damping}}\big)
  \underbrace{\frac{c}{c+\cb}}_{\sim\text{dimmer switch}} \ .
\end{align}\end{linenomath}
The single dimensionless parameter
$\gam$ depends on the geometry of the concentration filament
and chemotaxis parameters:
$\gam$ decreases for longer and thinner filaments,
while it increases with a rate of chemotactic re-orientation, see \SIref{sec:chemotactic-navigation} for details.
To leading order, the effective equation of motion
\eqref{dimless-ODE-main} represents a damped harmonic oscillator.
The corresponding frequency and damping ratio match the damped oscillation
observed in simulations, see \FIG{filament-surfing} and
\FIGSUPP{damped-oscillation}. The strong gradient in the
cross-section of the filament causes sperm cells to navigate towards
the centerline of the filament. Yet, cells continuously pass through
this centerline due to their finite chemotactic turning rate and
consequently oscillate within the filament. The much weaker gradient
along the concentration filament in \eqref{filament-main} damps this
oscillation when sperm cells move towards the egg, and amplifies it
when they move away.

The threshold $\cb$ of sensory adaption limits chemotaxis to the part
of the filament with concentration at least of the order of $\cb$.
This defines a cross-sectional area $A(x)$, where $c(\vec{r})\geq\cb$,
as well as circumference $S(x)$, at each centerline position $x$ of
the filament. We decompose the search for the egg into an \emph{outer
  search}, i.e., finding the concentration filament, and an
\emph{inner search}, i.e., surfing along the filament, see
\SIref{sec:effective-volume}. For the outer search, we introduce
the flux $j_{\text{out}}$ of sperm cells arriving at the surface of the
concentration filament and assume that $j_{\text{out}}$ is
approximately independent of the position $x$ along the filament.
Given that the egg has to be found within the exposure time $\tmax$,
we also introduce the outer search time
$t_{\text{out}}(x,\tmax)<\tmax$ available to arrive at the filament at
$x$ as specified below. For the inner search, using the effective
equation of motion, we compute the probability
$p_{\text{in}}(x,\tmax)$ that a sperm cell entering the filament at
position $x$ reaches the egg within time $\tmax$. We also compute the
conditional mean surfing time $t_{\text{in}}(x,\tmax)$, i.e., the
average time successful sperm cells require to reach the egg after entering
the filament at $x$. Correspondingly, we set the time for the outer
search as $t_{\text{out}}(x,\tmax) = \tmax-t_{\text{in}}(x,\tmax)$ for
$p_{\text{in}}>0$ (and $t_{\text{out}}=0$ for $p_{\text{in}}=0$). With
these prerequisites, we can formulate a general formula for the
sperm-egg \successs{} $\su$ in the presence of shear flow
\begin{linenomath}\begin{equation}
  \label{eq:theory-main}
  \su \approx \Int_{-\Rmax}^{\Rmax} \id{x} p_{\text{in}}(x,\tmax)
  \Big[A(x) \rho_{\text{egg}} + S(x) j_{\text{out}}
  t_{\text{out}}(x,\tmax) \Big] \ .
\end{equation}\end{linenomath}
The first term approximates the contribution from sperm cells that are
initially within the filament. This contribution is negligible
compared to the second term for low $\rho_{\text{egg}}$ or large
$\tmax$. The second term quantifies the contribution from sperm cells
that successfully find the concentration filament and surf along it to
the egg. The flux $j_{\text{out}}$ can be determined either from a fit
to full simulations or approximated as
$j_{\text{out}} = \rho_{\text{egg}} \vs/4$ by treating sperm cells
outside the filament as ballistic swimmers with speed $\vs$, see
\SIref{sec:effective-volume}, both of which gives similar
results. Moreover, the approximation of a ballistic swimming path
outside of the filament is reasonable, as the persistence length of
sperm swimming paths in the absence of chemoattractant cues was
estimated as $3 - 25~\text{mm}$~\cite{Fri2008b}, which is much greater
than the diameter of concentration filaments.

Note that for the chosen parameters, the volume
$V_{\text{tot}} = \int_{-\infty}^{\infty} \id{x} A(x)$ of the filament
(and its surface area $\int_{-\infty}^{\infty} \id{x} S(x)$) increases
monotonically with \shearrate{} \shear{}. Hence, the optimal
$\shearopt$ is not explained by a flow-dependent `chemotactic volume'
$V_{\text{tot}}$. Instead, the optimum emerges from two effects
related to filament surfing, which reduce $p_{\text{in}}$ and
$t_{\text{out}}$ in \eqref{theory-main} at high $\shear$: First, when
the filament is too thin at the entry point $x$ to enable the first
oscillation, the sperm cells simply pass through the filament, which
corresponds to low or vanishing probability $p_{\text{in}}$. Second,
if the time required to surf from the entry point $x$ to the egg is
too long, which corresponds to low or vanishing $t_{\text{out}}$, the
sperm cells will not reach the egg during the exposure time $\tmax$.
Higher \shearrate{}s generate longer and thinner filaments, which
aggravates both effects.

Comparison of full simulations and the theoretical prediction
\eqref{theory-main} shows good agreement, see \FIG{suc}. This
agreement strongly suggests that the optimal \shearrate{} $\shearopt$
originates from two competing effects: Higher shear flow spreads the
chemoattractant faster, which facilitates sperm navigation to the egg,
but results in longer and thinner filaments, which impairs chemotactic
filament surfing. The value of the optimal shear rate $\shearopt$
could be adjusted to a different value by re-scaling the biological
parameters that involve a time-scale such as the diffusion coefficient
and release rate of chemoattractant or the swimming and chemotactic
re-orientation speed of sperm cells, see
Sec.~\ref{sec:chemotactic-navigation} .

According to our theory, the presence of an optimal flow strength is a
generic feature at low egg densities and relatively long exposure
times. Amplitude and position of the peak of the sperm-egg-\successs{}
$\su(\shearopt)$ depend on chosen parameters. Our theory allows to
compute $\su(\shear)$ for any given set of parameters and thus the
parameter-dependency of the optimal \shearrate{} $\shearopt$ can be
explored. A numerical parameter study is presented in
\SIref{sec:parameter-variation}, which demonstrates the robustness of
the existence of an optimal \shearrate{} under parameter variation. In
short, a higher egg density $\rho_{\text{egg}}$ and longer exposure
time $\tmax$ increase the absolute amplitude $\suc(\shearopt)$ of
this peak, while $\shearopt$ stays almost constant. A high sensitivity
threshold $\cb$ of chemotactic signaling, which is formally analogous
to a high background concentration of chemoattractant, reduces the
relative amplitude $\su(\shearopt)/\su(\shear=0)$ of the
peak. Significantly shorter exposure time $\tmax$ or higher egg
density $\rho_{\text{egg}}$ reduce $p_{\text{in}}$ by effectively
cutting off the outer parts of the filament.
Note that the optimal \shearrate{} $\shearopt$ is slightly smaller in
simulations, compared to the theory. Inspection of simulated
trajectories suggest that this is due to sperm cells, which miss the
egg at least once while surfing along the filament, which increases
the mean surfing time $t_{\text{in}}$.

\subsection*{Comparison with experiments}

Previous experiments measured the fraction of fertilized eggs $\suc$
for an exposure time $\tmax$. This fraction directly relates to the
\successs{} $\su$ by fertilization
kinetics~\cite{VogCziChaWol1982,LevSewChi1991} when the respective
densities of sperm and egg cells, $\rho_{\text{sperm}}$ and
$\rho_{\text{egg}}$, are known
\begin{linenomath}\begin{align}
\label{eq:fertilization-kinetics}
  \suc(\tmax) = 1 -
  \exp\leri{-\sucfert\su(\tmax)\frac{\rho_{\text{sperm}}}{\rho_{\text{egg}}}}
  \ .
\end{align}\end{linenomath}
The fertilizability \sucfert{} is the probability that a
sperm-egg-encounter results in successful fertilization. Note that a
local maximum of the \successs{} $\su$ at some optimal \shearrate{}
$\shearopt$ automatically gives a local maximum of the \success{}
$\suc$. In particular, the density of sperm only alters the absolute
value of $\suc$ across all \shearrate{}s but not the existence and
value of an optimal \shearrate{} $\shearopt$.

\subsubsection*{Moderate shear} \label{sec:ZimRif2011}

\begin{figure}[t]
\includegraphics{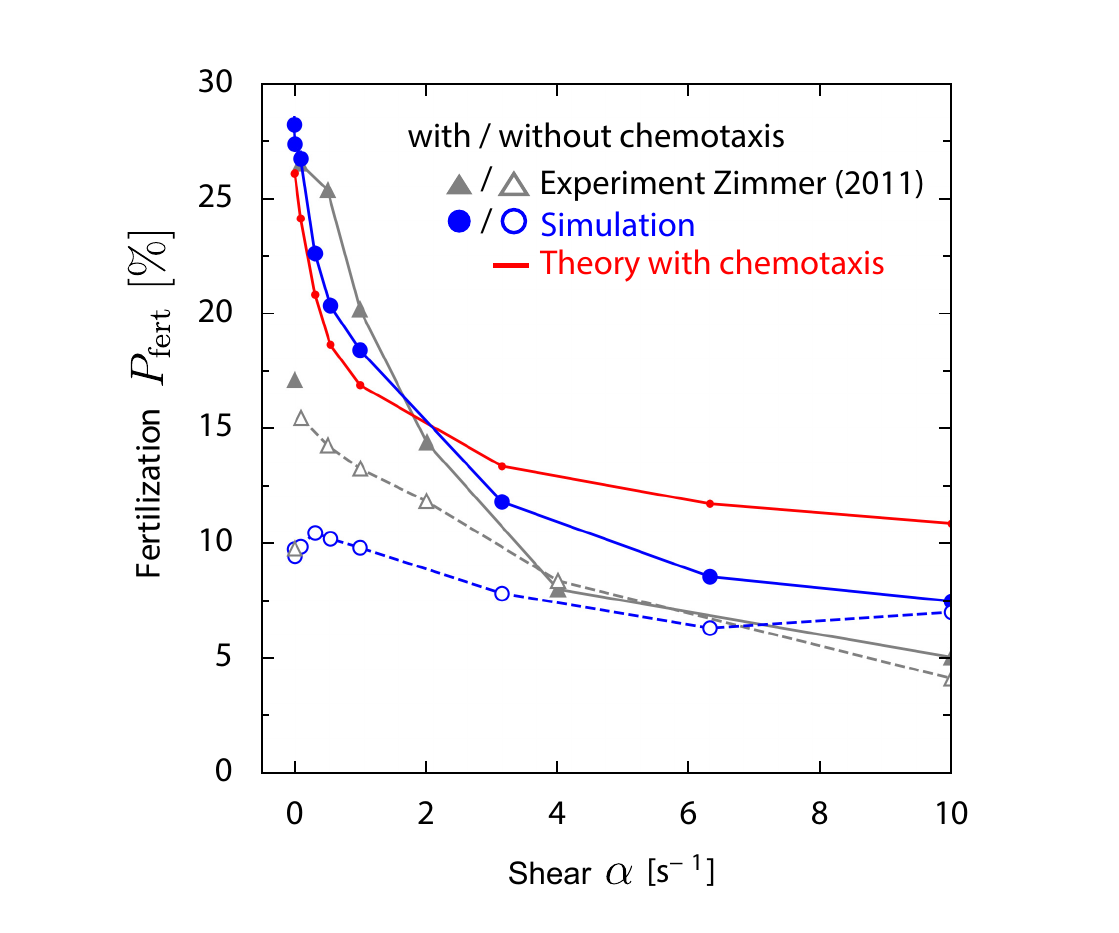}
\caption{\label{fig:fig5c} \textbf{Comparison to experiment at moderate
    \shearrate{}s and short exposure time.} Fertilization probability
  $\suc(\shear)$ that an egg becomes fertilized as function of
  external \shearrate{} $\shear$ from previous experiments with red
  abalone \emph{H. rufescens} gametes in a Taylor-Couette chamber
  (filled gray triangles: with chemotaxis, open gray triangles:
  inhibited chemotaxis; for $\shear = 0~\text{s}^{-1}$ a different
  experimental protocol was used)~\cite{ZimRif2011} and our
  corresponding simulations (filled blue circles: with chemotaxis,
  open blue circles: without chemotaxis, mean $\pm$ SD). We find
  reasonable agreement using a single fit parameter, fertilizability
  $\sucfert \approx 60 \%$, which characterizes the fraction of
  sperm-egg encounters that result in successful fertilization, see
  \eqref{fertilization-kinetics}. From the experimental protocol, we
  estimate a background concentration $c_{\text{bg}}\sim 4~\text{nM}$
  of chemoattractant. While our theory of filament surfing does not
  directly apply due to this high background concentration, a
  near-field estimate (red line) yields a similar decay of \success{}
  as function of \shearrate{} $\shear$. The single fit parameter of
  the theory,
  $j_{\text{out}} = 4.8\cdot10^3~\text{m}^{-2}\text{s}^{-1}$, is again
  consistent with the limit
  $j_{\text{out}} = \rho_{\text{egg}} \vs/4 =
  7.5\cdot10^3~\text{m}^{-2}\text{s}^{-1}$
  of a ballistic swimmer with random initial conditions (note the the
  higher value of $j_{\text{out}}$ compared to \FIG{suc} due
  to higher egg density). For simplicity, simulations do not account
  for co-rotation of sperm cells, see \FIGSUPP{fig5c-coro} for results
  with co-rotation.}
\end{figure}
\clearpage

In a previous experiment by Zimmer and Riffell, fertilization was
studied for red abalone \emph{H. rufescens} in a Taylor-Couette
chamber for moderate \shearrate{} \shear{}, mimicking flow conditions
in their natural spawning habitat~\cite{RifZim2007,ZimRif2011}. The
measured \success{} decreased with increasing $\shear>0$, both for
normal chemotaxis and a case of chemically inhibited chemotaxis, see
\FIG{fig5c} for a reproduction of the original
data~(\cite{ZimRif2011}, Fig.~5c). At low \shearrate{}, the measured
\success{} is twice as high with chemotaxis than without, while there
was little difference at high \shearrate{}s. This suggests that the
performance of sperm chemotaxis is reduced at high \shearrate{}s. We
performed simulations of sperm chemotaxis in external flow, using
parameters that match the specific experimental setup of
\cite{RifZim2007,ZimRif2011}, see \SIref{sec:constants}.
Specifically, the time span between preparation of the egg suspension
and the actual fertilization experiment results in a background
concentration of chemotattractant, which we estimate as
$c_{\text{bg}}\sim 4~\text{nM}$, i.e., several orders of magnitude
larger than the threshold of sensory adaption $\cb$, and account for
in the simulations. We compare results of these simulations and the
experiments, using fertilizability $\sucfert$ as single fit parameter,
see \FIG{fig5c}. We find good agreement for the case with
normal chemotaxis, and reasonable agreement for the case of inhibited
chemotaxis (potentially due to residual chemotaxis in the latter
case). An exception is the data point at $\shear=0~\text{s}^{-1}$. In
fact, a different experimental protocol was used for this data point,
corresponding to different initial mixing of sperm and egg cells,
which is not modeled in the simulations. In \FIG{fig5c}, we
neglected co-rotation of sperm cells for simplicity. We find similar
results if we account for co-rotation, except for the highest
\shearrate{}s, where \success{} is reduced, see
\FIGSUPP{fig5c-coro}. For simplicity, a \shearrate{} dependent chemokinesis as
suggested by \cite{RifZim2007,ZimRif2011}, i.e., regulation of
sperm swimming speed, is not included in the model, as preliminary
simulations suggest that this changes results only slightly. In our
comparison, we focused on the case of low sperm density
considered in \cite{RifZim2007,ZimRif2011}, thereby avoiding
confounding effects of sperm-sperm interactions and reduced
fertilization rates due to polyspermy at high sperm
densities~\cite{Sty1998,MilAnd2003}.

The absence of an optimal \shearrate{} $\shearopt$ is caused by the
high background concentration $c_{\text{bg}}$ in the experiment: Due to
$c_{\text{bg}}$, the part of the filament with sufficiently high
concentration $c(\vec{r}) \gtrsim \cb + c_{\text{bg}}$ is situated
only in the vicinity of the egg and has an approximately spherical
shape. While our far-field theory of filament surfing does not
directly apply to this special near-field case, a simple estimate for
$p_{\text{in}}$ and $t_{\text{out}}$ assuming straight sperm
trajectories aligned with the local concentration gradient inside the
plume, see \SIref{sec:effective-volume}, yields a similar decay of
\success{}, see \FIG{fig5c}. The fitted flux of sperm cells into the
concentration plume
$j_{\text{out}} = 4.8\cdot10^3~\text{m}^{-2}\text{s}^{-1}$ is
consistent with the limit
$j_{\text{out}} = \rho_{\text{egg}} \vs/4 =
7.5\cdot10^3~\text{m}^{-2}\text{s}^{-1}$ for a ballistic swimmer. This
validates our interpretation of chemotaxis in external shear as a
two-stage search, consisting of blind random search for a chemotactic
volume and subsequent navigation inside this volume.
We emphasize that
the high background concentration of chemoattractant, which we
reconstruct for these experiments, has a strong effect on the
fertilization dynamics. Such high background concentrations are
unlikely to be encountered in natural habitats, where eggs are
spawned and consequently diluted in the open water.

\subsubsection*{Strong flows} \label{sec:MeaDen1995}

\begin{figure}[hb!]
\includegraphics{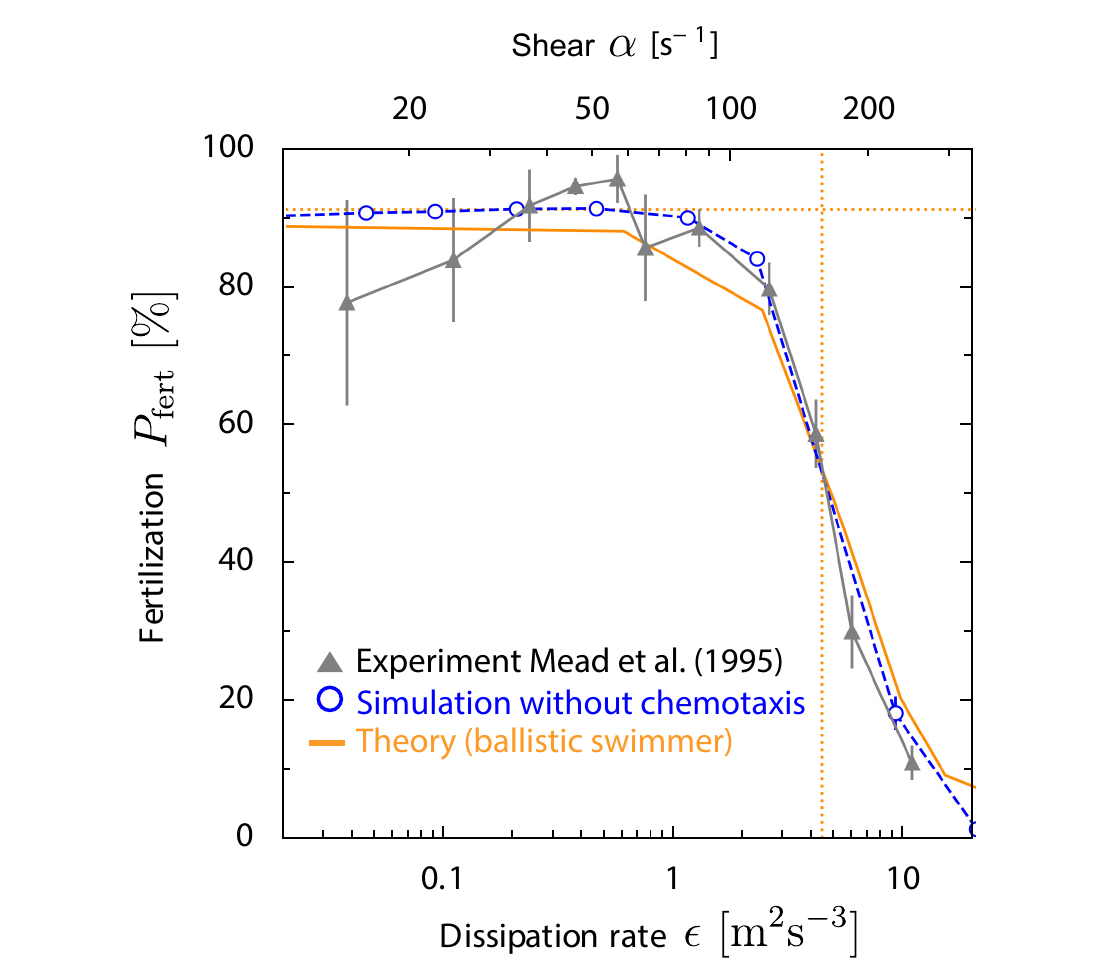}
\caption{\label{fig:fig5} \textbf{Fertilization in strong flows and high
    egg density.} Previous measurements of \success{} $\suc(\eps)$ for
  sea urchin \emph{S. purpuratus} at strong turbulence, characterized
  by density-normalized dissipation rate $\eps$ (filled gray
  triangles) \cite{MeaDen1995,Gay2008} and our corresponding
  simulations $\suc(\shear)$ as function of \shearrate{} $\shear$
  (open blue circles, mean $\pm$ SD) match well, using a single fit
  parameter $a=0.075$ that relates dissipation rate $\eps$ and typical
  \shearrate{} $\shear$ (using the known relationship
  $\shear(\eps) = a
  \sqrt{\eps/\nu}$~\cite{LazMan1989,JumTroBosKar2009}). Both
  simulation and experiment are well captured by a minimal theory of a
  ballistic swimmer in simple shear flow (red), see
  \SIref{sec:stokes-flow}. Fertilization probability $\suc$
  rapidly drops above a characteristic flow strength
  $\shear>100~\text{s}^{-1}$, which is consistent with a scale
  estimate $\shear=2 \pi \vs/(0.1 \Regg)$ (vertical dotted line). At
  these high \shearrate{}s, active swimming becomes negligible
  compared to convection. The case of low \shearrate{}s is well
  described by the limit case of a ballistic swimmer in the absence of
  flow $\shear = 0~\text{s}^{-1}$ (dotted horizontal line,
  \eqref{fertilization-kinetics} with $\su(t)=1-\exp(-\rate t)$ and
  rate $\rate=\pi \Regg^2 \vs \rho_{\text{egg}}$). The fertilizability
  $\sucfert = 10 \%$ is obtained from an independent
  experiment~\cite{MeaDen1995}, see \FIGSUPP{fig4}. From the
  experimental protocol, we estimate a high background concentration
  $c_{\text{bg}} = 500-4000~\text{nM}$ of chemoattractant, which
  renders sperm chemotaxis ineffective. Corresponding results for
  simulations with co-rotation are shown in \FIGSUPP{fig5-coro}.}
\end{figure}

Mead and Denny studied fertilization in the sea urchin \emph{S.
  purpuratus} in turbulent flow, mimicking physiological conditions in
the oceanic surf zone~\cite{MeaDen1995, DenNelMea2002,
  Gay2008}. The measured fertilization probability slightly increased
as function of turbulence strength, quantified in terms of local
dissipation rate $\eps$, and decreased rapidly at larger dissipation
rate $\eps>1~\text{m}^2\text{s}^{-3}$, see \FIG{fig5} for a
reproduction of the original data (taken from Fig.~3 of
\cite{Gay2008}, representing a re-calibration of data from Fig.~5
of \cite{MeaDen1995}). We determined \success{} \suc{} in simple
shear flow from simulations, using parameters that match the specific
experimental setup, see \SIref{sec:constants}. For the
experiments by Mead and Denny, we estimate a high background
concentration of chemoattractant $c_{\text{bg}} = 500-4000~\text{nM}$,
which renders sperm chemotaxis ineffective, which is thus neglected in
the simulations. Fully developed turbulence is characterized by a
spectrum of local \shearrate{}s, with a characteristic \shearrate{}
\shear{} related to the dissipation rate by
$\shear(\eps) = a \sqrt{\eps/\nu}$ with proportionality factor
$a$~\cite{LazMan1989,JumTroBosKar2009}. In the
simulations, we assume a simple shear flow $\vf=\shear y \, \e_x$, and
determine $a=0.075$ by a single-parameter fit, see
\FIG{fig5}. For sake of simplicity, co-rotation of sperm cells is
neglected. Results with co-rotation are qualitatively very similar,
yet the \success{} $\suc$ drops at a smaller \shearrate{} \shear{} and
thus yields a smaller fit parameter $a=0.023$, see
\FIGSUPP{fig5-coro}. Note that these fits for $a$ are smaller
than values commonly used in the literature
$a \sim 0.15 - 1.8$~\cite{Kol1941,Kol1962,LazMan1989}. Nevertheless,
our minimal model already reproduces the experimentally observed
characteristic drop in \success{} $\suc(\eps)$ at high flow rates,
implying that this is a robust, general feature.

We can capture the functional dependence of the \success{} \suc{}
observed in both experiment and simulations by a minimal theory of a
ballistic swimmer in simple shear flow, see \FIG{fig5} and
\SIref{sec:stokes-flow}. In particular, for small \shearrate{}
\shear{}, \suc{} is close to the asymptotic limit $\suc(\shear=0)$ of
a ballistic swimmer without flow. The drop of \suc{} at strong flow
can be estimated from a simple scaling argument: At high \shearrate{}
$\shear\geq \vs/\Regg$, the active swimming of sperm cells is
negligible compared to the external flow, except in the direct
vicinity of the egg. This vicinity is set by a characteristic distance
$\delta \sim 0.1 \, \Regg$ from the egg, up to which the flux of sperm cells
is elevated (due to the geometry of the streamlines around the egg).
To reach the egg, these sperm cells have to traverse a distance $\sim\delta$
within the typical time $t_\delta\sim 2\pi / \shear$ that
corresponding streamlines spend in the vicinity of the egg (time for
half rotation of the egg). Thus, the characteristic flow strength at
which $\suc$ drops can be estimated as
$\shear \sim 2\pi \vs / \delta$, see \FIG{fig5}.

For \FIG{fig5}, we obtain the fertilizability
$\sucfert \approx 10\%$ from an independent experiment in the absence
of flow~\cite{MeaDen1995}, which is well described by the
fertilization kinetics, \eqref{fertilization-kinetics}, see
\FIGSUPP{fig4}. This $\sucfert$ is larger than a value
$\sucfert = 3.4 \%$ previously reported for sea urchin \emph{S.
  franciscanus} ~\cite{LevSewChi1991,VogCziChaWol1982}.
However, these previous experiments were conducted at much higher
sperm densities, where sperm-sperm interactions and
polyspermy~\cite{Sty1998,MilAnd2003} may reduce the \success{}.
The estimated fertilizability for sea urchin is smaller
than our estimate for red abalone $\sucfert=60\%$, which is expected
due to the jelly coat of sea urchin eggs: For red
abalone, sperm cells are considered to arrive directly on the egg surface,
whereas for sea urchin, sperm cells are considered to arrive at a jelly coat
surrounding the egg, which sperm cells have to penetrate before
fertilization.

\section*{Discussion}

\begin{figure}[hb!]
\includegraphics[width=0.9\textwidth]{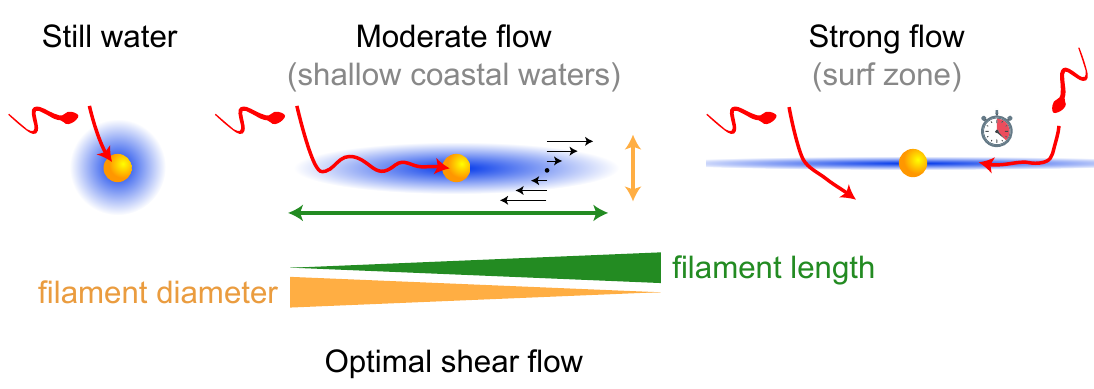}
\caption{\label{fig:discussion} \textbf{Proposed mechanism explaining
    optimal flow strength for sperm chemotaxis.} Egg cells (yellow)
  release signaling molecules (blue) that guide sperm cells of marine
  species with external fertilization (red). External flows (black
  arrows) stretch concentration gradients into millimeter-long
  filaments. If sperm cells encounter such filaments, they can
  ``surf'' by chemotaxis towards the egg. In strong flows, however,
  sperm cells may fail to follow the filament after encounter, because
  the effective diameter of filaments is too small. Additionally, in
  very long filaments, sperm cell may not reach the egg within the
  sperm-egg exposure time (which is set by the lifetime of the
  smallest eddies for turbulent flow). Thus, it is not the total
  volume of the chemoattractant plume that determines fertilization
  success, but the geometric shape of filaments. The competition
  between increasing filament length, which favors sperm-egg
  encounters, and decreasing filament diameter, which jeopardizes
  filament surfing, sets an optimal flow strength that maximizes
  sperm-egg encounters. The optimal flow strength predicted by our
  theory matches physiological flow strengths in typical habitats.}
\end{figure}

We presented a general theory of sperm chemotaxis at small-scale turbulence,
using marine sperm chemotaxis in physiological shear flow as
prototypical example. We predict that sperm chemotaxis performs better
in physiological flows as compared to conditions of still water. Our
theory provides a novel phenomenological description of concentration
filaments shaped by external flow, and describes how sperm cells surf
along these filaments in terms of damped oscillations. Extensive
simulations show that fertilization success becomes maximal at an
optimal flow strength. We explain the existence of this optimal flow
strength as the result of a competition between a faster built-up of
concentration gradients in the presence of flow, and the
disadvantageous distortion of concentration fields into increasingly
thinner concentration filaments at increased flow rates, see also
Fig.~\ref{fig:discussion}. The optimal flow rate predicted by our
theory matches typical flow strengths in typical spawning habitats in
shallow coastal
waters~\cite{LazMan1989,MeaDen1995,RifZim2007,JumTroBosKar2009,CriZim2014}.
The maximal sperm-egg encounter probability at the optimal flow rate
depends strongly on egg density and sperm-egg exposure time, see
parameter study in the Supporting Information. In contrast, the optimal flow rate
$\shearopt$ as predicted by our theory is largely independent of egg
density, sperm-egg exposure time, and other parameters.

In our simulations, we considered a constant sperm-egg exposure time,
independent of flow strength, following the experimental protocol from
\cite{MeaDen1995,Gay2008,RifZim2007,ZimRif2011}. For fully developed
turbulence in natural habitats, the life time of the smallest eddies
sets an effective sperm-egg exposure time, as the turn-over of
small-scale eddies resets local concentration fields surrounding the
egg. As a consequence, the sperm-egg exposure time decreases with
increasing flow strength (approximately,
$\tmax \approx 25/\alpha \sim \varepsilon^{-1/2}$, using the decay
time scale of corresponding Burger vortices, see
Sec.~\ref{sec:constants} ). This
provides yet a third effect that reduces the success of sperm
chemotaxis in strong turbulence.
While the steady shear flow is a
minimal model, we expect more realistic turbulence simulations to
yield similar results at the relevant scales based on simulations with
an unsteady shear flow, as shown in \FIG{filament-surfing}A: although, the
shear rate is rather large $\shear = 0.17~\text{s}^{-1}$, i.e., almost
twice as large as the shear rate at the optimum, concentration filaments
are only slightly bent by the rotational diffusion of the shear axis in unsteady shear flow,
and we consistently find that sperm cells surf along slender concentration filaments, as exemplified in \FIG{filament-surfing}A.

Our numerical simulations quantitatively account for previous
fertilization experiments in Taylor-Couette chambers. These
experiments impressively demonstrated the reduction of fertilization
success at high flow rates and hinted at the existence of an optimal
flow rate, which motivated our theoretical study. Our theoretical
analysis highlights two subtleties in the interpretation of these
early experiments, i.e., a high background concentration of
chemoattractant and possibly insufficient mixing of sperm and egg
cells in the absence of flow, both of which can confound an existing
optimum. Nonetheless, we are indebted to this pioneering work and can
now predict conditions, under which an optimal flow strength is
expected. This can aid the rational design of future experiments.
While a direct experimental observation of filament surfing is
pending, recent 3D tracking experiments of sea urchin sperm cells
navigating in axially symmetric chemoattractant landscapes gave
intriguing anecdotal evidence how these cells first found the
centerline of these concentration filaments and subsequently moved
parallel to this centerline \cite{JikAlvFriWilPasColPicRenBreKau2015}.
While our numerical simulations consider a specific mechanism of sperm
chemotaxis along helical paths, our analytical theory is more general
and applies in particular to any chemotaxis strategy for which the
swimming direction gradually aligns to the local gradient direction.
This suggests that the presence of an optimal flow strength could be a
general phenomenon.

We expect that our findings of two-stage chemotactic search,
comprising finding a filament and subsequent surfing along this
filament, could be also relevant for foraging of bacteria and
plankton: Nutrient patches are stirred by turbulent flow into networks
of thin filaments, in which these organisms have to navigate for
optimal uptake. Finding sinking marine snow from whose surface
nutrients dissolve bears resemblance to finding egg cells which
release chemoattractant. These organisms play an important role for
oceanic
ecosystems~\cite{LucBerMit1999,LocPed2009,JumTroBosKar2009,TaySto2012,Sto2012,KioSai1995,BreLalWaaWilMaz2018,LomKosKio2013,BruCarHeiHagLevSto2020}.
While our theory addresses the experimentally more accessible model
system of external fertilization as employed by marine invertebrates
\cite{Mil1985}, chemotaxis in external flows is relevant also for
internal fertilization, where sperm cells navigate complex
environments \cite{SuaPac2006,GafGadSmiBlaKir2011}, likely guided by
both chemotaxis \cite{EisGio2006} and rheotaxis
\cite{MikCla2013,KanDunBlaGol2014,MarFuPowSto2012}.
We emphasize that rheotaxis and chemotaxis in the presence of external flow as considered here
rely on different physical mechanisms,
despite formal similarities, such as active swimming upstream an external flow.
While rheotaxis relies on the co-rotation of active swimmers,
we found co-rotation to be dispensable for successful chemotactic navigation,
with upstream swimming arising solely from chemotactic alignment to concentration filaments shaped by flow.

More generally, we characterized
sperm chemotaxis in external flow as a combination of random
exploration, followed by local gradient ascent, which corroborates a
general paradigm for cellular and animal search behavior
\cite{HeiCarBruStoLev2016}. The minimalistic information processing
capabilities of sperm cells (comparable to that of a single neuron
\cite{Kau2012}) can inspire biomimetic navigation strategies for artificial
microswimmers with limited information processing capabilities
intended for navigation in dynamic and disordered environments
\cite{LanYamRyaYamSanKat2019,XuMedMagSchHebSch2018}.

\section*{Methods and Materials}
\label{sec:methods-materials}

The \successs{} \su{} is computed numerically by simulating sperm
trajectories $\vec{r}(t)$ in the presence of both a concentration
field $c(\vec{r})$ of chemoattractant and an external fluid flow field
$\vf(\vec{r})$ according to equations of motion for $\vec{r}(t)$, see
\SIref{sec:eq-motion}. These equations extend a previous,
experimentally confirmed theory of sperm chemotaxis along helical
paths~\cite{FriJue2007,JikAlvFriWilPasColPicRenBreKau2015} by
incorporating convection and co-rotation of cells by the external
flow. For co-rotation, we employ Jeffery equation for prolate
spheroids~\cite{Jef1922,PedKes1992} by assigning sperm cells an effective aspect ratio
$\geofactor=5$. For the shear rates considered here, the effect of
external flow on sperm flagellar beat patterns is
negligible~\cite{KliRulWagFri2016}. Each sperm cell is simulated for
an exposure time $\tmax$, which is set by protocol of the
corresponding experiment, or until it hits the surface of the egg.

As external flow, we assume a simple shear flow around a
freely-rotating spherical egg, see \SIref{sec:stokes-flow}. Throughout,
we consider the co-moving frame of the egg allowing us to assume that
the egg is at the origin $\vec{r}=\vec{0}$. The concentration field is
established by diffusion and convection from the egg releasing
chemoattractant at a constant rate. We consider the reference case of
a static concentration field corresponding to a chemoattractant
release time equal to exposure time $\tmax$. Note that the exposure
time $\tmax$ may also be estimated by the
decay time scale of a Burger vortex $\tmax \sim 25/\shear$, see \SIref{sec:numerical-simulation}. To account for an
ensemble of eggs at density $\rho_{\text{egg}}$, we consider a single
egg with radius $\Regg$ at the origin $\vec{r}=\vec{0}$ and a
spherical domain with radius
$\Rmax=\leri{4 \pi \rho_{\text{egg}}/3}^{-1/3}$ and appropriate
periodic boundary conditions: Initially, sperm cell positions
$\vec{r}$ ($\Regg\leq\norm{\vec{r}}\leq\Rmax$) and directions of the
helix axis $\h$ are uniformly distributed, representing the
distribution after initial turbulent mixing of egg and sperm cells. If
sperm cells leave the simulation domain, they re-enter with random new
initial conditions $\vec{r}$ and $\h$ with $\norm{\vec{r}}=\Rmax$,
whose distribution $P_b(\vec{r},\h)$ is defined by the theoretical
in-flux of cells due to active swimming and convection
\begin{linenomath}\begin{align}
\label{eq:boundary}
  P_b\leri{\vec{r},\h} \sim - p_{\text{sperm}}\leri{\vec{r},\h}
  \lerii{\leri{\vf\leri{\vec{r}} + \vs\h}\cdot\e_r\leri{\vec{r}}}
\end{align}\end{linenomath}
with uniform and isotropic distribution of sperm cells
$p_{\text{sperm}}\leri{\vec{r},\h}$. In principle, co-rotation of
non-spherical particles by shear flow leads to a non-uniform
distribution of directions $\h$, see analytic solutions in
\SIref{sec:jeff-eq-anal}, but the effect on simulation results
is negligible.

Parameters for \FIG{suc} were chosen to closely match
conditions of \emph{A. punctuala} sea urchin in their natural spawning
habitat at low egg density $\rho_{\text{egg}}$ and relatively long
exposure times $\tmax$. Parameters for Figs~\ref{fig:fig5c} and
\ref{fig:fig5} are chosen to match the experiments by Zimmer and
Riffell~\cite{RifZim2007, ZimRif2011} and Mead and
Denny~\cite{MeaDen1995}, respectively. For further details on
simulations and extensive discussion of parameters used for each scenario, see
\SIref{sec:constants}, \SIref{sec:numerical-simulation}. Finally,
error bars for simulation results represent simple standard deviation
(SD) of the corresponding binomial distribution. Error bars are
smaller than symbol sizes in some cases.

\section*{Acknowledgments}
  We are grateful for discussions with L.~Alvarez, M.~ Wilczek,
  M.~W.~Denny, and J.~Riffell. 

  \nocite{MikMor2004, FriJue2009, Rod1840, FraAcr1968, Elr1962,
    ZoeSta2012, ZoeSta2013, SloAlcSecStoFer2020, Mea1996,
    AlvDaiFriKasGrePasKau2012, HorBenBis2018, Fri2008, Picetal2014,
    HatKam1997, WebYou2015, BerPur1977,
    KrolaFri2020} 

\newpage
\clearpage

\section*{Supporting Information text}

\renewcommand{\theequation}{S\arabic{equation}}
\setcounter{section}{1} 
\setcounter{subsection}{0} 
\renewcommand{\thesubsection}{\Alph{subsection}}
\setcounter{equation}{0}  
\renewcommand{\thefigure}{\Alph{figure}}
\setcounter{figure}{0}  
\renewcommand{\thetable}{\Alph{table}}
\setcounter{table}{0}  

\begin{figure}[h!]
\hspace*{-1.5cm}\includegraphics{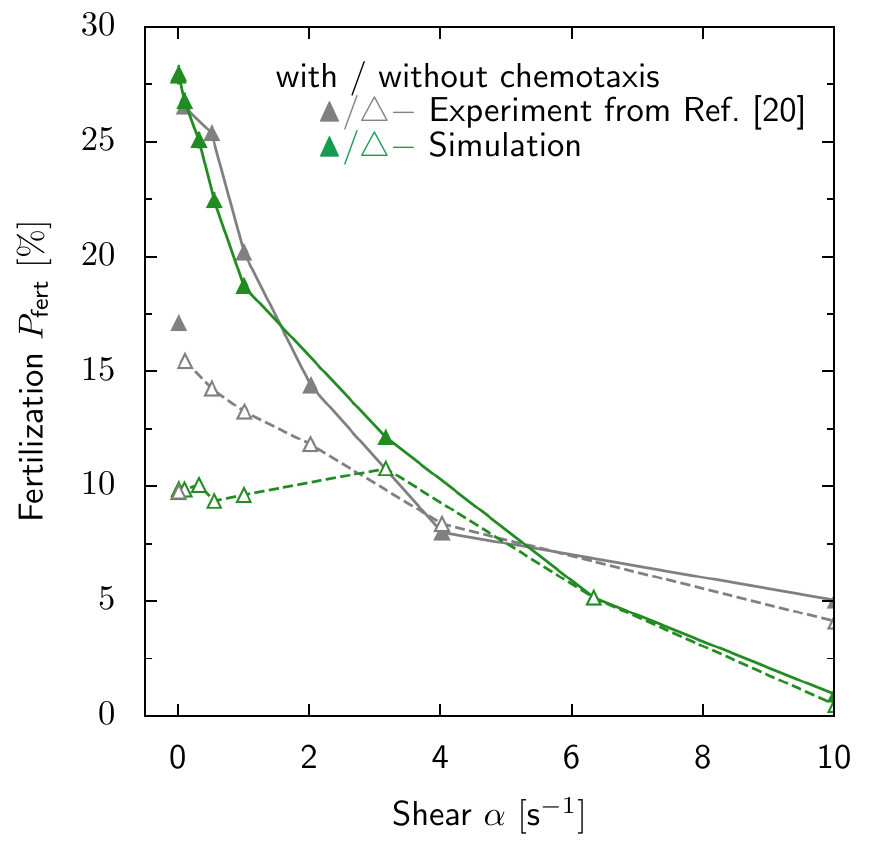}
\caption{
\label{fig:fig5c-coro}
\textbf{Fertilization probability with co-rotation.}
  Fertilization
  probability $\suc(\shear)$ as in \FIG{fig5c} but with
  simulations taking into account co-rotation (green triangles):
  Previous experimental data~(\cite{ZimRif2011}, Fig.~5c) for red
  abalone \emph{H. rufescens} with and without chemotaxis (filled gray
  triangles: with chemotaxis, open gray triangles: inhibited
  chemotaxis) and our corresponding simulations (filled green
  triangles: with chemotaxis, open green triangles: without
  chemotaxis; mean $\pm$ SD), using fertilizability $\sucfert = 60 \%$
  in \eqref{fertilization-kinetics} as single fit parameter.
  Experiment and simulation again agree reasonably except for the data
  point without flow $\shear = 0~\text{s}^{-1}$, which corresponds to
  a different experimental protocol. While the simulations with
  co-rotation overestimate the reduction of $\suc$ at high
  \shearrate{} $\shear>6~\text{s}^{-1}$, these high \shearrate{}s are
  less relevant for the spawning habitat of \emph{H. rufescens}.
}
\end{figure}

\begin{figure}[h!]
\hspace*{-1.5cm}\includegraphics{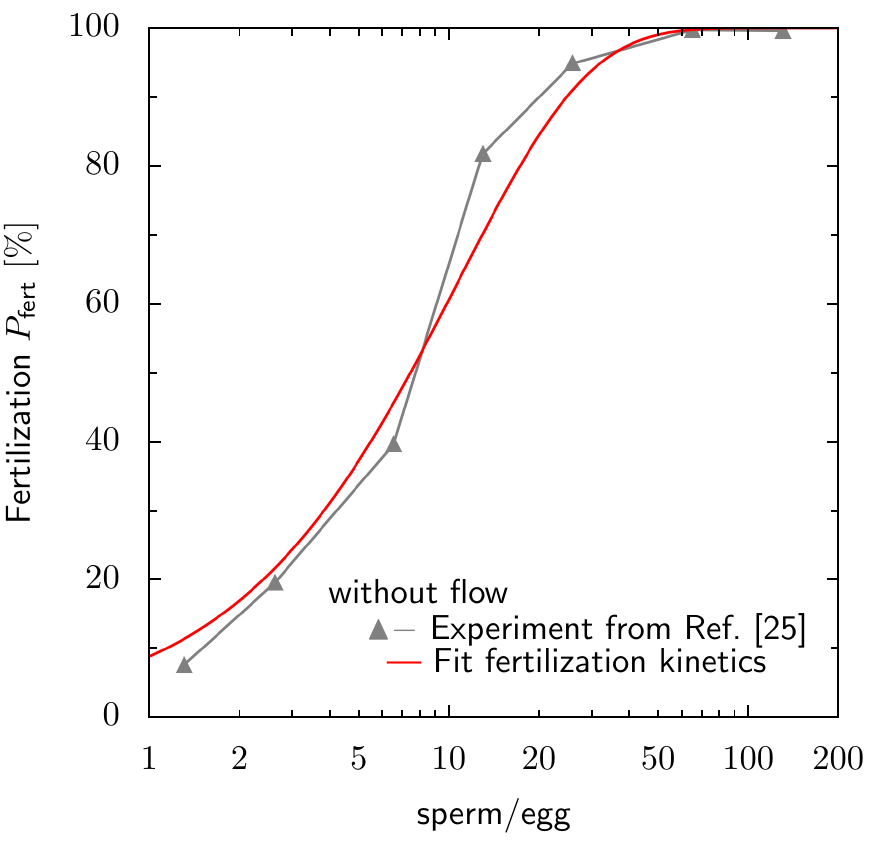}
\caption{
\label{fig:fig4}
  \textbf{Calibration of fertilizability without flow ($\shear = 0$).}
  Fertilization probability $\suc$ as function
  of the ratio $\rho_{\text{sperm}}/\rho_{\text{egg}}$ of sperm and
  egg density in the absence of flow $\shear = 0~\text{s}^{-1}$ from
  experiments with sea urchin \emph{S. purpuratus}~(\cite{MeaDen1995},
  Fig.~4)
  (filled gray triangles) and
  fit of fertilization kinetics \eqref{fertilization-kinetics} (red).
  From the fit, we obtain $\sucfert\su\approx 9 \%$ for the product of
  fertilizability $\sucfert$ and \successs{} $\su$. Assuming a
  ballistic swimmer that is captured at the egg surface
  (\eqref{fertilization-kinetics} with
  $\su(\tmax)=1-\exp(-\rate \tmax)$ and rate
  $\rate = \pi \Regg^2 \vs \rho_{\text{egg}} = 0.02~\text{s}^{-1}$),
  we find $\sucfert \approx 10 \%$ for exposure time
  $\tmax = 120~\text{s}$ and
  $\rho_{\text{egg}} = 1.5\cdot10^4~\text{ml}^{-1}$. This value
  $\sucfert$ is used in \FIG{fig5}, \FIGSUPP{fig5-coro}.
}
\end{figure}

\begin{figure}[h!]
\hspace*{-1.5cm}\includegraphics{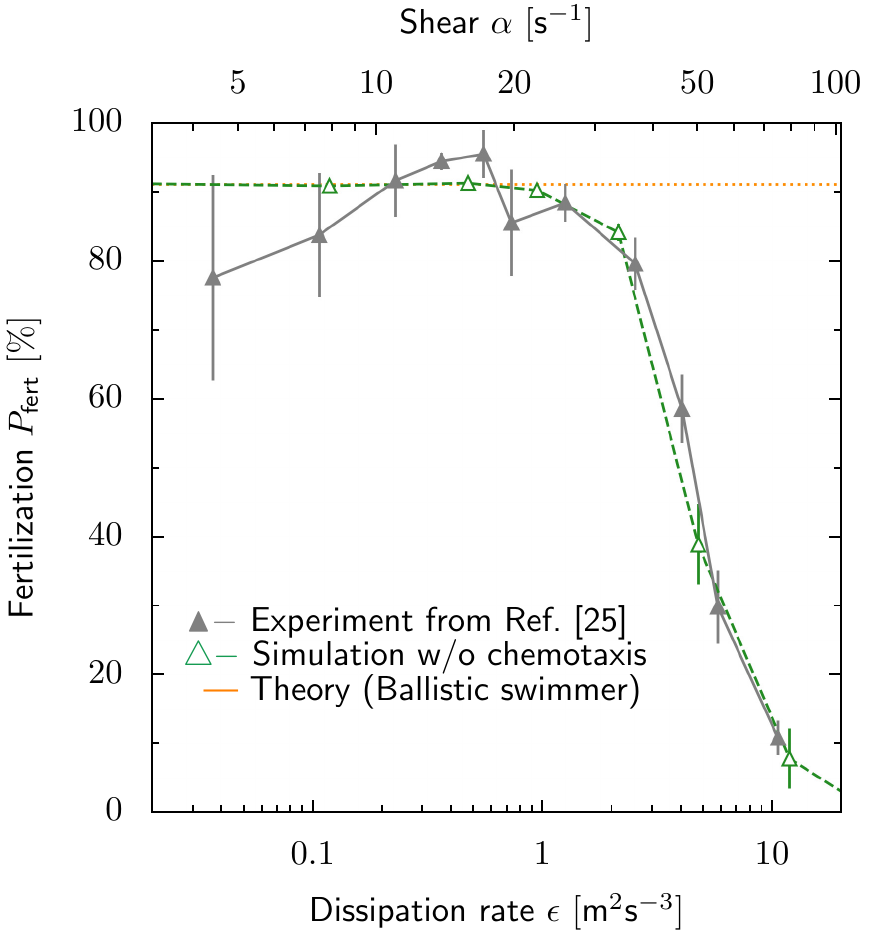}
\caption{
\label{fig:fig5-coro}
  \textbf{Fertilization probability with co-rotation.}
  Fertilization
  probability $\suc(\shear)$ as in \FIG{fig5} but with
  simulations taking into account co-rotation of sperm cells (open
  green triangles): Previous measurements of \success{} $\suc(\eps)$
  for sea urchin \emph{S. purpuratus} at strong turbulence,
  characterized by density-normalized dissipation rate $\eps$ (filled
  gray triangles) \cite{MeaDen1995,Gay2008} and our corresponding
  simulations $\suc(\shear)$ as function of \shearrate{} $\shear$
  (open green triangles, mean $\pm$ SD) match well, using a single fit
  parameter $a=0.023$ that relates dissipation rate $\eps$ and typical
  \shearrate{} $\shear$ (with the known relationship
  $\shear(\eps) = a
  \sqrt{\eps/\nu}$~\cite{LazMan1989,JumTroBosKar2009}). Analogous to
  \FIG{fig5}, the case of low \shearrate{}s is well described
  by the limit case of a ballistic swimmer in the absence of flow
  $\shear = 0~\text{s}^{-1}$ (dotted horizontal line,
  \eqref{fertilization-kinetics} with $\su(t)=1-\exp(-\rate t)$ and
  rate $\rate=\pi \Regg^2 \vs \rho_{\text{egg}}$). The fertilizability
  $\sucfert = 10 \%$ is obtained from an independent
  experiment~\cite{MeaDen1995}, see \FIGSUPP{fig4}. From the
  experimental protocol, we estimate a high background concentration
  $c_{\text{bg}} = 500-4000~\text{nM}$ of chemoattractant, which
  renders sperm chemotaxis
  ineffective.
}
\end{figure}

\subsection{Shear flow around freely-rotating egg and minimal case of
  ballistic swimmer} \label{sec:stokes-flow}

For all simulations (except \FIG{filament-surfing}A), we use
a simple shear flow $\shear y \, \e_x$ as idealized paradigm for
small-scale turbulence. At the relevant \shearrate{}s $\shear$ and
typical egg radii $\Regg\sim 100~\mu\text{m}$, the Reynolds number
$\re = \shear \Regg^2 / \nu \leq 0.1$ is sufficiently small to justify
the use of the analytical Stokes equation for viscous flow
$\vf(\vec{r})$. Throughout, we consider the co-moving frame of the egg
allowing us to assume that the egg is at the origin $\vec{r}=\vec{0}$.
We introduce dimensionless coordinates
$\hat{\vec{r}} = \vec{r} / \Regg$ and the dimensionless flow field
$\vfb\leri{\vec{\rb}} = \frac{2 \vf\leri{\vec{r}}}{\shear \Regg}$. The
components of this flow field read~(\cite{MikMor2004}, Eq.~(12))
\begin{linenomath}\begin{equation}
\label{eq:stokes-flow}
\begin{aligned}
  \vb{}_{\text{ext},x} &= 2 \yb &- \yb \leri{\lerii{1 + \hat{\Omega}}
    \rb^{-3} + \rb^{-5}} &- 5 \xb^2 \yb
  \leri{\rb^{-5}-\rb^{-7}} \\
  \vb{}_{\text{ext},y} &= &\phantom{+} \xb \leri{\lerii{1 +
      \hat{\Omega}}
    \rb^{-3}-\rb^{-5}} &- 5 \xb \yb^2 \leri{\rb^{-5}-\rb^{-7}} \\
  \vb{}_{\text{ext},z} &= & &- 5 \xb \yb \zb \leri{\rb^{-5}-\rb^{-7}}
\end{aligned}
\end{equation}\end{linenomath}
where no-slip boundary conditions on the surface $\norm{\vec{\rb}}=1$
of the freely-rotating spherical egg are assumed. The egg rotates
according to the undisturbed flow vorticity with the dimensionless
rotation rate $\hat{\Omega} = - 1$, corresponding to an rotation of
the egg with angular velocity $\vec{\Omega} = -\frac{\shear}{2} \e_z$.

It is instructive to consider a ballistic swimmer in the above flow
field $\vf$ as a reference for the analysis of more complicated cases,
such as swimmers performing chemotaxis. For instance, without flow or
chemotaxis, sperm cells are considered to swim along a straight helix with
helix radius $r_0$ much smaller than the egg radius. These sperm
trajectories are well approximated by a ballistic swimmer moving along
the helix axis $\h$ with net swimming speed $\vs$. If the ballistic
swimmers and the target eggs (with density $\rho_{\text{egg}}$) are
uniformly distributed, the steady-state rate $\rate$ at which a
swimmer hits an egg is given by
$\rate = \pi (\Regg+r_0)^2 \vs \rho_{\text{egg}} \approx \pi \Regg^2
\vs \rho_{\text{egg}}$.
If ballistic swimmers become trapped at the egg on encounter, this
corresponds to the \successs{} $\su(t)=1-\exp(-\rate t)$ (and
\success{} $\suc$ according to fertilization kinetics, see
\eqref{fertilization-kinetics}). If ballistic swimmers are
additionally convected by an external fluid flow field $\vf$, we can
characterize $\rate$ (and thus $\su$ and $\suc$) in terms of an
universal curve: We introduce the dimensionless parameter
$\k = \frac{\shear \Regg}{2 \vs}$, which compares \shearrate{} to net
swimming speed. The combined velocity field of active swimming and
fluid flow is now
\begin{linenomath}\begin{align}
  \vf\leri{\vec{r}} + \vs\h = \vs
  \leri{\k \vfb\leri{\vec{\rb}} + \h} =
  \vs\ub\leri{\vec{\rb}, \k, \h} \ .
\end{align}\end{linenomath}
Note that without co-rotation, $\h$ does not change. Thus, for any
$\h$, all possible velocity fields $\ub$ are given by a single
one-parameter family parametrized by $\k$. For each of these fields,
the dimensionless rate of swimmers $\hat{\rate}\leri{\k,\h}$ reaching
the egg from $\norm{\vec{\rb}}\gg 1$ specifies the actual rate $\rate$ for any
set of parameters $\shear, \Regg, \vs, \rho_{\text{egg}}$ with the
same parameter $\k$ by
\begin{linenomath}\begin{align}
  \rate\leri{\shear, \Regg, \vs, \h} =
  \hat{\rate}\leri{\k,\h} \Regg^2 \vs \rho_{\text{egg}} \ .
\end{align}\end{linenomath}
We obtain a universal curve for $\rate$ by computing
$\hat{\rate}(\k,\h)$ numerically for all $\k$ and $\h$ and average
$\hat{\rate}(\k) = \leriv{\hat{\rate}(\k,\h)}_{\h}$ over all
directions $\h$, see \FIG{fig5} for corresponding $\suc$. A
prominent feature of the universal rate is that it vanishes at large
\shearrate{}s $\hat{\rate}\leri{\k\to \infty} \to 0$. In the absence
of flow $\shear=0$, we have $\hat{\rate}\leri{\k = 0} = \pi$.

We compute the universal rate $\hat{\rate}$ efficiently by integrating
a uniform grid of initial conditions on the surface of the egg, with
$\norm{\vec{\rb}}=1$ at $\hat{t}=0$, backwards in time according to
the velocity field $\ub$. Each initial condition is integrated until
it either returns to the egg $\norm{\vec{\rb}}(\hat{t})=1$ (fail) or
leaves the outer boundaries
$\vec{\rb}(\hat{t}) = \hat{r}_{\text{max}}$ (success) with
$\hat{r}_{\text{max}} \gg 1$. As the flow is volume conserving, the
results are independent of the choice of the outer boundary
$\hat{r}_{\text{max}}$, as long as $\hat{r}_{\text{max}}$ is
sufficiently large to ensure the absence of closed orbits beyond it.
We choose $\hat{r}_{\text{max}} = 4$ as numerics show that the in- and
outflow on this sphere differs only by $4 \%$ between the Stokes flow
around the freely-rotating sphere and the undisturbed simple shear
flow, for which it is known no closed orbits exist. Based on the
intersections with the outer boundary, the flow reaching the egg is
interpolated. This is done for a grid of swim directions $\h$. For
efficiency, we exploit the symmetries of the Stokes flow
$\vfb\leri{\xb,\yb,\zb} \cdot \leri{h_x \e_x + h_y \e_y + h_z \e_z} =
\vfb\leri{\xb,\yb,-\zb} \cdot \leri{h_x \e_x + h_y \e_y - h_z \e_z} =
\vfb\leri{-\xb,-\yb,\zb} \cdot \leri{-h_x \e_x - h_y \e_y + h_z
  \e_z}$;
thus, it is sufficient to consider $h_z \geq 0$ and $h_y\geq 0$,
respectively.

\subsection{Equations of motion for navigating sperm cells}
\label{sec:eq-motion}

We simulate the swimming path $\vec{r}(t)$ of a sperm cell in a
concentration field $c(\vec{r})$ of chemoattractant in the presence of
an external fluid flow field $\vf(\vec{r})$. For this, we extend a
previous theory of chemotaxis of marine sperm cells along helical
paths~\cite{FriJue2007,FriJue2009,AlvFriGomKau2014,JikAlvFriWilPasColPicRenBreKau2015}
by incorporating convection and co-rotation by flow: The sperm cell is
described in terms of the time-dependent center position $\vec{r}(t)$,
averaged over one flagellar beat cycle, and the set of ortho-normal
vectors $\e_1(t) , \e_2(t), \e_3(t)$ of the co-moving coordinate
frame, where the vector $\e_1(t)$ points in the direction of active
swimming with speed $\vo$. The equations of motion read
\begin{linenomath}\begin{equation}
\label{eq:eq-motion}
\begin{aligned}
  \dt{\vec{r}} &= \vo \e_1 + \vf(\vec{r}(t)) \ , \\
  \dt{\e}_i &= \leri{\vec{\Omega}_{\text{h}} +
    \vec{\Omega}_{\text{f}}} \times \e_i \qquad i = 1,2,3 \ ,
\end{aligned}
\end{equation}\end{linenomath}
The two angular velocities, $\vec{\Omega}_{\text{h}}$ and
$\vec{\Omega}_{\text{f}}$, describe the rotation of the coordinate
frame due to helical chemotaxis and external flow, respectively. For
\eqref{eq-motion} a constant swim speed is assumed and motility noise
is neglected; the persistence length of sperm swimming paths in the
absence of chemoattractant cues was estimated as
$3 - 25~\text{mm}$~\cite{Fri2008b} which validates this assumption.
Note that \eqref{eq-motion} is also valid for time-dependent
concentration and flow fields.
Note further that the quantitative
comparison of the two angular velocities in \eqref{eq-motion} already
suggests that the rotation due to external flow is negligible, as the
rate of change due to external flow
$\Omega_{\text{f}} \sim \shear = 0 - 1~\text{s}^{-1}$ (see
\eqref{jeff-eq}) is always smaller than due to the helical motion
$\Omega_{\text{h}} \sim \tor_0 \vo = 3 - 13~\text{s}^{-1}$ (see
\eqref{helical-omega} and parameters in Table~\ref{tab:constants}).

Without external flow or chemotaxis, cells swim along a helical path
with constant path curvature $\curv(t)=\curv_0$ and torsion
$\tor(t)=\tor_0$. The angular velocity $\vec{\Omega}_{\text{h}}$ is
defined by the Frenet-Serret equations
\begin{linenomath}\begin{align}
  \label{eq:helical-omega}
  \vec{\Omega}_{\text{h}}(t) = \vo \lerii{\tor(t) \e_1(t) + \curv(t)
  \e_3(t)} \ ,
\end{align}\end{linenomath}
where the coordinate frame $\e_1 , \e_2, \e_3$ corresponds to the
Frenet-Serret frame of $\vec{r}(t)$, i.e., tangent, normal and
bi-normal vector. During chemotactic steering, sperm cells dynamically
regulate curvature $\curv(t)$ and torsion $\tor(t)$ of active swimming
according to the output $a(t)$ of a chemotactic signaling system
\begin{linenomath}\begin{equation}
\begin{aligned}
  \curv(t) &= \curv_0 - \psperm \curv_0 (a-1) \ , \\
  \tor(t) &= \tor_0 + \psperm \tor_0 (a-1) \ .
\end{aligned}
\end{equation}\end{linenomath}
Here, the sensori-motor gain factor $\psperm$ characterizes the
amplitude of chemotactic steering responses. The chemotactic signaling
system takes as input the local concentration $c(\vec{r}(t))$ at the
position of the cell
\begin{linenomath}\begin{equation}
\label{eq:eq-output}
\begin{aligned}
  \mu \dt{a} &= p\lerii{\cb + c(\vec{r}(t))} - a \ , \\
  \mu \dt{p} &= p\leri{1-a} \ .
\end{aligned}
\end{equation}\end{linenomath}
This minimal signaling system comprises sensory adaption with
sensitivity threshold $\cb$ and relaxation with time scale $\mu$ to a
rest state $a=1$ for any constant stimulus $c(\vec{r}(t)) = c_0$. The
variable $p$ describes an dynamic sensitivity which is regulated down
when the stimulus is high, or regulated up when the stimulus is low (a
loose analogy would be that $p$ corresponds to the opening of our
eye's pupils as adaption to brightness). In principle, $p$ and $a$
could have different time-scales~\cite{FriJue2007}. However, equal
time-scales automatically ensure that the phase-lag between
small-amplitude oscillations of the input signal $c(\vec{r}(t))$ and
resulting oscillation of the output signal $a(t)$ attains the value
$\pi/2$ optimal for helical chemotaxis~\cite{FriJue2009}. This special
case is sufficient for the purpose of a minimal model. The gain factor
$\psperm$ sets the rate of chemotactic steering. While $\psperm$ could
depend on the chemotactic signal by a feedback mechanism
~\cite{KroMaeLanBaiFri2018}, we assume here a constant gain factor
$\psperm=5$ for simplicity. The values of all parameters are listed
and discussed in \SI{sec:constants}.

We approximate the angular velocity $\vec{\Omega}_{\text{f}}$ for
co-rotation by external flow using the Jeffery equation for a small
prolate spheroid with major axis along $\e_1$~\cite{Jef1922,PedKes1992}
\begin{linenomath}\begin{equation}
  \label{eq:jeff-eq}
\begin{aligned}
  \vec{\Omega}_{\text{f}}(\vec{r}) &= \frac{1}{2} \vec{\omega}(\vec{r})  + \G \e_1 \times
                          \lerii{\mat{E}(\vec{r}) \cdot \e_1} \ , \\
  \vec{\omega}(\vec{r}) &= \Nabla \times \vf(\vec{r}) \ , \\
  \mat{E}(\vec{r}) &=  \frac{1}{2} \lerii{\Nabla \otimes \vf(\vec{r}) +\leri{\Nabla \otimes \vf(\vec{r})}^T}
\end{aligned}
\end{equation}\end{linenomath}
with the flow vorticity $\vec{\omega}$, the strain rate tensor
$\mat{E}$, and a geometric factor
$\G = \frac{\geofactor^2-1}{\geofactor^2+1}$, which depends on the aspect
ratio $\geofactor \geq 1$ of major to minor axis of the spheroid. Together
with \eqref{eq-motion}, \eqref{jeff-eq} describes the cell rotation
due the flow, i.e., The first term in the first line of
\eqref{jeff-eq} describes rotation of a spherical body due to flow
vorticity and the second term the correction for non-sperical bodies
that can be approximated as spheroids. For a swimming sperm cell, we
take the swim direction $\e_1$ as effective major axis, and employ an
effective aspect ratio, $\geofactor=5$, reflecting the ratio of the length
of the flagellum and a typical beat amplitude~\cite{RifZim2007}. Note
that in general instead of $\e_1$, the major axis could be any
co-moving vector.

We numerically integrate the equations of motion, i.e.,
Eqs.~(\ref{eq:eq-motion},\ref{eq:eq-output}), using an Euler scheme
with fixed small time step $dt$. For efficient computation, Rodrigues
rotation formula~\cite{Rod1840} with respect to the co-moving coordinate frame is
used to integrate $\e_1 , \e_2, \e_3$, resulting in faster computation
compared to the algorithm used in \cite{KroMaeLanBaiFri2018}.

\subsection{Analysis of concentration filaments}
\label{sec:filament-anal}

Turbulent flows cause turbulent mixing of diffusing chemicals and
generate filamentous concentration fields. As a minimal model, we simplify the turbulent
flow and the filamentous concentration field by the case of a simple
shear flow. We consider a spherical egg located at the origin
$\vec{r}=\vec{0}$ releasing chemoattractant with diffusion coefficient
$\D$ at a constant rate $\relrate$ in the presence of shear flow
$\vf(\vec{r})$ given by \eqref{stokes-flow}. We compute the
time-dependent concentration field $c(\vec{r},t)$ of chemoattractant
numerically using Lagrangian particle tracking, see
\SI{sec:constants}. We empirically find that the far-field at
distances $r \gg \Regg$ is well approximated by a generic profile, see
\FIG{filament-surfing}B for illustration,
\begin{linenomath}\begin{align}
\label{eq:typical-filament}
  c(\vec{r},t) = c_0 \exp\leri{-\lam \abs{x}}
  \exp\leri{-\frac{\leri{y-y_0}^2/\ratio^2 + z^2}{2 \sig^2}} \ ,
\end{align}\end{linenomath}
which describes a concentration filament with
time-dependent parameters $c_0(t)$,
$\lam(t)$, $\ratio$,
as well as
time- and position-dependent variance and midline profiles $\sig(x,t)$
and $y_0(x,t)$, respectively.
This formula for the
concentration filament is consistent with results obtained using the
analytic solution for an instantaneous point source in a shear flow
$\shear y \, \e_x$, see below. We present and discuss scaling laws for
the parameters in the following. While these dependencies are not
explicitly required for our theory, they demonstrate the universality
of our theory. Finally, we use these scaling laws to quantify how the
filaments become longer and thinner with increasing \shearrate{}
$\shear$.

From numerical simulations, we empirically find the following scaling
laws of the parameters from \eqref{typical-filament}
\begin{linenomath}\begin{align}
\label{eq:cfil_scaling}
  \lam\leri{t} &\sim
                 t^{-\delta_\lam} \ , &\delta_\lam& = 1.5 \text{ -- } 1.6 \\
  c_0\leri{t} &\sim t^{-\delta_{c_0}} \ , &\delta_{c_0}& = 1.6 \text{ -- } 1.8 \tag{\ref{eq:cfil_scaling}b} \\
  \sig^2\leri{x,t} &= 2 \D \, t_0\leri{x,t} \ ,
                                      &t_0&\sim t \tag{\ref{eq:cfil_scaling}c} \\
  t_0\leri{x,t} &\omit\rlap{$\ = \ptO\leri{t} + \ptI\leri{t}
                        \abs{x} + \ptII\leri{t} \abs{x}^2,$} \tag{\ref{eq:cfil_scaling}d} \\
  y_0\leri{x,t} &= \sgn\leri{x} \pyO\leri{t} + \pyI{} t^{-\delta_{y_0}}x \ , &\delta_{y_0}& = 0.9 \text{ -- } 1.0 \tag{\ref{eq:cfil_scaling}e} \\
  \pyI{}  &= (1 \text{ -- } 1.4)
              \shear^{-\delta_{y_0}} \ , \qquad \qquad &\ratio& = 0.5 \text{ -- } 0.6 \tag{\ref{eq:cfil_scaling}f}
\end{align}\end{linenomath}
where all parameters except $\ptII$ are positive. Note that in a
turbulent flow the time $t$ in which the filament is formed may scale
with the Kolmogorov time $t \sim \tkol$; in this case $\sigma$ would
scale as the Batchelor length $\sigma \sim \sqrt{D \tkol}$. We also
found power-law dependencies for the coefficients $\ptO(t)$,
$\ptI(t)$, and $\ptII(t)$. The factor $\ratio$ appears to be constant
for sufficiently large $t$. These numerical observations become
plausible by analysis of a point source in shear flow. The
Fokker-Planck equation for this case can be written in dimensionless
form
\begin{linenomath}\begin{align}
  \partial_{t} c = -\shear y\, \partial_{x} c
  + \D \bigtriangleup c
  \,+\, \relrate\,\delta(\vec{r})
\Rightarrow
  \partial_{\check{t}} \check{c} =
  - \check{y}\, \partial_{\check{x}} \check{c}
  + \bigtriangleup \check{c}
  \,+\,\delta(\check{\vec{r}})
\end{align}\end{linenomath}
by using the Batchelor scale $\sqrt{\D \tkol} \sim \sqrt{D/\shear}$ to
re-scale to dimensionless coordinates
\begin{linenomath}\begin{align}
\label{eq:rescaling}
  \check{x} = x \sqrt{\frac{\shear}{\D}} \ , \quad \check{y} = y \sqrt{\frac{\shear}{\D}} \ , \quad \check{t} = t
  \shear \ , \quad \check{c} = c
  \sqrt{\frac{\D^3}{\shear}} \frac{1}{\relrate}
\end{align}\end{linenomath}
with \shearrate{} $\shear$, and release rate $\relrate$ of the source
(i.e. $\iiint_{-\infty}^{\infty} \mathrm d^3 r \, c(\vec{r},t) =
\relrate\, t$).
Consequently, the solution $\check{c}\leri{\vec{\check{r}},\check{t}}$
of this equation can be re-scaled to the solution $c(\vec{r},t)$ for
any set of parameters $\shear, \D, \relrate$. For the above form of
the far-field of the filament, this implies that the parameters
$\delta_\lam$, $\delta_{c_0}$, $\pyI$ and $\ptII$ are universal as they are invariant
under the re-scaling \eqref{rescaling}. The analytical solution for the
dimensionless concentration $\check{c}$
reads~(\cite{FraAcr1968}, Eq.~(18))
\begin{linenomath}\begin{align}
\label{eq:anal-point}
  \check{c}\leri{\vec{\check{r}},\check{t}} = \Int_0^{\check{t}} \id{\check{s}}
  \check{G}(\vec{\check{r}},\check{s})
\end{align}\end{linenomath}
with Greens function $\check{G}$, i.e., the solution for an
instantaneous source at the origin~(\cite{Elr1962}, Eq.~(26))
\begin{linenomath}\begin{align}
  \check{G}(\vec{\check{r}}, \check{t}) = \frac{\exp\lerii{-\frac{\leri{\check{x}-\frac{1}{2}\check{y}\check{t}}^2}{4\check{t}\leri{1+\frac{1}{12}\check{t}^2}}
  - \frac{\check{y}^2+\check{z}^2}{4\check{t}}}}{\leri{4\pi
  \check{t}}^{\frac{3}{2}} \sqrt{1+\frac{1}{12}\check{t}^2}} \ .
\end{align}\end{linenomath}
While the integral \eqref{anal-point} cannot be solved analytically,
it explains the empirical scaling for the parameters in
\eqref{typical-filament} heuristically: It is reasonable to assume
that for any $\check{x}$, the parameter $\check{y}_0(\check{t})$ is
close to the point $\check{y}_{\text{max}}$ of the maximal
concentration of $\check{G}(\check{\vec{r}}, \check{t})$. From
$\partial_{\check{y}}
\check{G}(\check{x},\check{y},\check{z}=0,\check{t})|_{\check{y}=\check{y}_{\text{max}}}
= 0$, it follows (for $\check{t} > \sqrt{3}$)
\begin{linenomath}\begin{align}
  \check{y}_0(\check{x},\check{t}) \approx \check{y}_{\text{max}}(\check{x},\check{t})=\frac{3 \check{t}
  \check{x}}{2(\check{t}^2+3)} \Rightarrow \pyI{} \approx \frac{3}{2}\check{t}^{-1}
\end{align}\end{linenomath}
in accordance with the fitted power-law.

The power law $c_0(t) \sim t^{-\frac{3}{2}}$, as suggested by
numerics, is plausible since
$\check{G}\leri{\vec{0},\check{s}\gg 1} \sim
\check{s}^{-\frac{5}{2}}$,
which implies
$\check{c}\leri{\vec{0},\check{t}\gg 1} \sim \int_0^{\check{t}}
\id{\check{s}} \check{s}^{-\frac{5}{2}} \sim
\check{t}^{-\frac{3}{2}}$.

We introduce the concentration $\check{c}_{\text{max}}$ at the
centerline of the filament
$\check{c}_{\text{max}}\leri{\check{x},\check{t}} =
\check{c}\leri{\check{x},\check{y}_0\leri{\check{x},\check{t}},\check{z}=0}$.
We make the ansatz
$\check{c}_{\text{max}}\leri{\check{x},\check{t}} =
\check{c}_0(\check{t}) \exp\leri{-\check{\lam}\leri{\check{t}}
  \abs{\check{x}}}$
and derive a power-law for $\check{k}(\check{t})$ in the following. We
expect that $\check{c}_{\text{max}}$ scales proportional to the summed
contributions of the Greens functions at the time-dependent
centerline, hence we estimate (assuming $\check{t}\gg 1$, we
approximate $\check{t}^2+3 \to \check{t}^2$,
$1+\check{t}^2/12 \to \check{t}^2/12$ in $\check{G}$)
\begin{linenomath}\begin{align}
\label{eq:cmax-approx}
  \check{c}_{\text{max}}\leri{\check{x},\check{t}} \sim \Int_0^{\check{t}} \id{\check{s}} \check{G}(\check{x},\check{y}_0(\check{x},\check{s}),\check{z}=0,\check{s})
  \sim \frac{\erfc{\leri{\sqrt{\frac{3}{4\check{t}^3}} \check{x}}}}{6\pi
  \check{x}} \ .
\end{align}\end{linenomath}
We are interested in the shape of the concentration filament up to a
maximal distance $\check{x}_{\text{max}}$ at which the concentration
at the centerline decayed to a fraction $\iota$ of $\check{c}_0$,
$\check{c}_{\text{max}}(\check{x}_{\text{max}},\check{t})=\iota
c_0(\check{t})$.
Any asymptotic tails beyond this distance will likely not be relevant
for chemotaxis. Since the decay of $\check{c}_{\text{max}}$ as
function of $\check{x}$ is dominated by the numerator in
\eqref{cmax-approx}, the distance $\check{x}_{\text{max}}$ has a
time-dependency
$\check{x}_{\text{max}}\leri{\check{t}} \sim \check{t}^{\frac{3}{2}}$
according to the argument of the complementary error-function $\erfc$.
Using \eqref{cmax-approx}, we estimate the time-dependency of
$\check{\lam}(\check{t})$ from
\begin{linenomath}\begin{align}
  \label{eq:k-approx}
  -\check{k}(\check{t}) \check{x} = \ln\leri{\frac{\check{c}_{\text{max}}(\check{x})}{\check{c}_0}} \sim
  \ln\leri{\frac{\erfc{a}}{a}} \sim \ln \erfc\leri{a} \sim -a \ ,
\end{align}\end{linenomath}
where we introduced
$a(\check{x},\check{t}) = \sqrt{3/4} \
\check{t}^{-\frac{3}{2}}\check{x}$.
The crucial point is that for
$0\leq \check{x}\leq\check{x}_{\text{max}}(\check{t})$, the variable
$a$ varies only in a finite interval $0\leq a \leq a_{\text{max}}$
with upper bound
$a_{\text{max}}=a(\check{x}_{\text{max}}(\check{t}),\check{t})\sim\check{t}^{-\frac{3}{2}}\check{x}_{\text{max}}(\check{t})$
independent of time $\check{t}$. This allows us to approximate
$\ln \erfc\leri{a}$ by its Taylor expansion for small $a\ll 1$ in the
last step of \eqref{k-approx}. We conclude
$\check{\lam}\leri{\check{t}} \sim
a_{\text{max}}/\check{x}_{\text{max}}(\check{t}) \sim
\check{t}^{-\frac{3}{2}}$, as suggested by numerics.

From the above considerations follows that for a constant exposure
time $\tmax$ the filaments become longer and thinner with increasing
\shearrate{} $\shear$: From \eqref{rescaling} follows for the
dimensionless exposure time $\check{t}_{\text{max}} = \tmax \shear$
and thus the exponent $\check{\lam}$ of the dimensionless version of
\eqref{typical-filament} scales with
$\check{\lam} \sim (\tmax\shear)^{-\frac{3}{2}}$, which implies
according to $\check{\lam}\check{x}=\lam x$ a scaling of the effective
decay length
\begin{linenomath}\begin{align}
  1/\lam \sim \shear \sqrt{\D \tmax^3} \ .
\end{align}\end{linenomath}
This means that for constant exposure time $\tmax$ the effective
length of the filament increases with \shearrate{} $\shear$.
Analogously, from $\check{\sigma} \sim \sqrt{\tmax \shear}$ and
$\check{c}_0 \sim \leri{\tmax\shear}^{-\frac{3}{2}}$ follows with the
re-scaling \eqref{rescaling} for the effective decay length $\sig$
away from the center of the filament and the base concentration $c_0$
\begin{linenomath}\begin{equation}
\begin{aligned}
  \sig &\sim \sqrt{\D \tmax} \\
  c_0 &\sim \frac{\relrate}{\shear \leri{\D \tmax}^{\frac{3}{2}}} \ .
\end{aligned}
\end{equation}\end{linenomath}
The combination of the effective decay length $\sig$ being independent
of $\shear$ and the base concentration $c_0$ decreasing with
increasing $\shear$ means that the effective width of the filament
decreases with increasing $\shear$.

\subsection{Chemotactic navigation within filament}
\label{sec:chemotactic-navigation}

We derive an effective equation of motion for chemotactic navigation
within a typical concentration filament. For simplicity, we initially
ignore interaction with the flow and assume that the motion is
effectively two-dimensional, i.e., in the $xy$-plane. Additionally, we
employ a two-dimensional version of \eqref{typical-filament} for the
concentration filament, setting $\ratio=1$,
\begin{linenomath}\begin{align}
  \label{eq:filament-2D}
  c(x,y,t) = c_0 \exp\leri{-\lam \abs{x}}
  \exp\leri{-\frac{(y-y_0)^2}{2 \sig^2}} \ .
\end{align}\end{linenomath}
We introduce the centerline $\vec{r}_h(t)=(x(t),y(t),0)$ of the
helical swimming path $\vec{r}(t)$, with $\dt{\vec{r}}_h = \vs \h$.
From a previously established equation for
$\vec{r}_h$~\cite{FriJue2007,FriJue2009}, we have
\begin{linenomath}\begin{equation}
\label{eq:chemotactic-navigation}
\begin{aligned}
  \dt{x} &= \vs\cos\leri{\phi} \ , \qquad \dt{y} = \vs\sin\leri{\phi}
  \ , \\
  \dt{\phi} &= - \bet \frac{\norm{\Nabla c}}{c + \cb} \sin \Psi \ ,
  \qquad \Psi = \sphericalangle \leri{\Nabla c,\h} \ ,
\end{aligned}
\end{equation}\end{linenomath}
describing the alignment of the helix axis $\h$ with the local
gradient $\Nabla c(\vec{r}_h(t))$ of a concentration field
$c(\vec{r})$. The first equation corresponds to ballistic motion along
the helix axis $\h(\phi) = \cos \phi \ \e_x + \sin \phi \ \e_y$ with
net swimming speed $\vs = \vo \tor_0 /\sqrt{\curv_0^2+\tor_0^2}$. The
second equation describes chemotactic turning of the orientation angle
$\phi$, where $\Psi$ denotes the angle enclosed by $\h$ and the local
gradient $\Nabla c(\vec{r}_\h(t))$. Here, $\cb$ denotes the adaption
threshold and $\bet$ the chemotactic turning speed,
$\bet = \psperm \vs \curv_0^2 / \leri{\curv_0^2+\tor_0^2}$, $\bet>0$,
with the gain factor $\psperm$ and helix parameters $\curv_0$,
$\tor_0$. We apply this general theory,
\eqref{chemotactic-navigation}, to the filamentous profile
\eqref{filament-2D} and obtain a single dimensionless ODE
\begin{linenomath}\begin{align}
\label{eq:dimless-ODE}
  \ddt{\y} = \big(\underbrace{-\dt{\x}^2\y}_{\sim\text{oscillator}} +
  \underbrace{\sign{\x} \ \gam
  \dt{\x}\dt{\y}}_{\sim\text{damping}}\big)
  \underbrace{\frac{c}{c+\cb}}_{\sim\text{dimmer switch}}
\end{align}\end{linenomath}
with $\dt{\x}^2+\dt{\y}^2 = 1$, $\dt{\x}\neq 0$ and a single
dimensionless parameter
\begin{linenomath}\begin{align}
  \label{eq:dimless-gam}
  \gam = \lam \sig \sqrt{\frac{\bet}{\vs}} = \lam \sig
  \sqrt{\frac{\psperm \curv_0^2}{\curv_0^2+\tor_0^2}} \ .
\end{align}\end{linenomath}
Here, we introduce a characteristic time-scale $\utau$,
\begin{linenomath}\begin{align}
\label{eq:time-scale}
  \utau = \sqrt{\frac{\sig}{\bet}\cdot\frac{\sig}{\vs}} \ ,
\end{align}\end{linenomath}
as well as re-scaled coordinates $\y(\t) = \leri{y(t)-y_0}/L$,
$\x(\t) = x(t)/L$, $L = \vs \tau$. Dots denote differentiation with
respect to re-scaled time $\t = t / \utau$, e.g.,
$\dt{\y} = \id{\y}/\id{\t}$. The time scale $\utau$ is the geometric
mean of a characteristic time-scale $\sig/\bet$ of chemotactic
steering and a typical time $\sig/\vs$ for traversing the
cross-sectional width $\sig$ of the filament if steering was absent.
We have an equation for $\x$ analogous to \eqref{dimless-ODE} (which
requires $\dt{\y} \neq 0$ and covers the case $\dt{\x} = 0$),
\begin{linenomath}\begin{align}
\label{eq:dimless-ODEx}
\ddt{\x} = \leri{\dt{\x}\dt{\y} \y - \sign{\x} \ \gam \dt{\y}^2}
  \frac{c}{c+\cb} \ .
\end{align}\end{linenomath}
The factor $c/(c+\cb)$ in the effective equations of motion,
Eqs.~(\ref{eq:dimless-ODE}, \ref{eq:dimless-ODEx}), represents a
`dimmer switch' that attenuates chemotactic navigation at low
concentration $c$. Thus, it is reasonable to define the filament as
the region where $c(\vec{r})\geq\cb$. In the following, we focus on
the dynamics within the filament and approximate $c/(c+\cb)\approx1$.

\begin{figure}[ht]
  \includegraphics{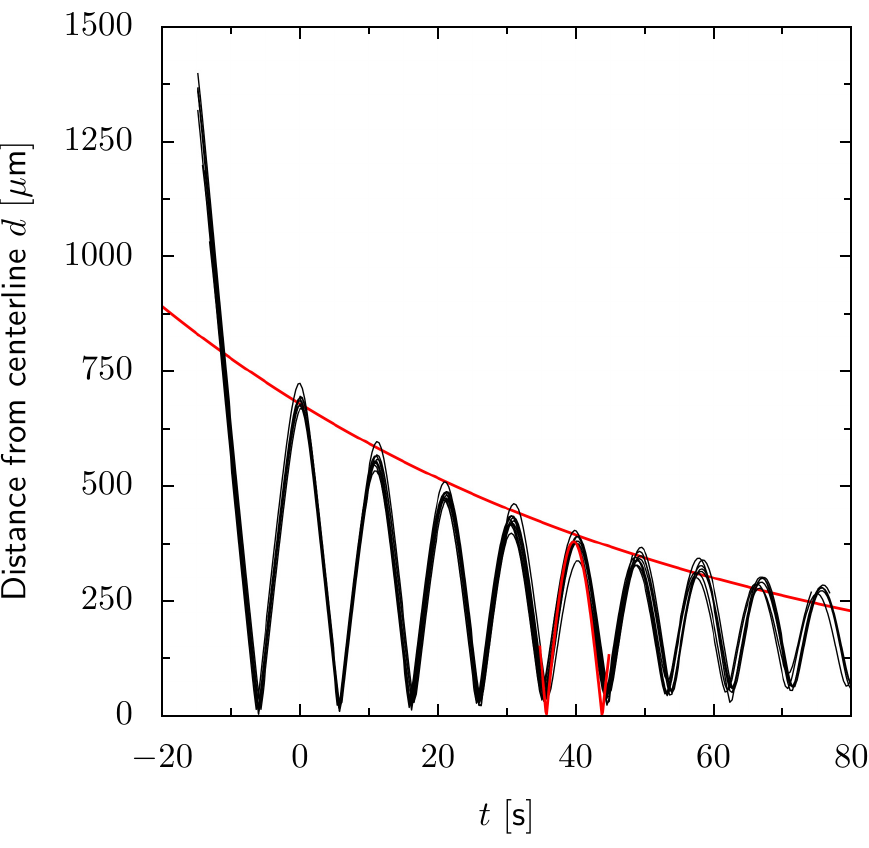}
\caption{\label{fig:damped-oscillation} \textbf{Surfing along filaments
    can be described as damped oscillation.} Distance
  $d = \sqrt{(y-y_0)^2+z^2}$ from the centerline of concentration
  filament \eqref{typical-filament} superimposed for $n=9$ sperm
  trajectories simulated according to \SI{sec:eq-motion}
  (black). Trajectories are shown after they entered the surface of
  the concentration filament, defined by $c(\vec{r}(t))=\cb$, and
  shifted in time to align the first oscillation peak at $t_0=0$.
  Remarkably, all trajectories display stereotypic oscillations that
  overlap perfectly, despite the fact that trajectories entered the
  filament at different $x$-positions and initial direction angles.
  The observed damped oscillation are well reproduced by a minimal
  analytical theory for the centerline of the helical swimming path,
  which predicts damping ratio and oscillation period (dashed red
  line, see \eqref{frequency-damping-ratio}). Parameters as in
  \FIG{filament-surfing}B, corresponding to \emph{A.
    punctuala}.}
\end{figure}

The effective equation of motion, \eqref{dimless-ODE}, describes a
damped, non-linear oscillator: The first term $\dt{\x}^2 \y$
originates from the perpendicular component
$\Nabla_\perp c = \leri{\e_y \cdot \Nabla c} \e_y$ of the
concentration filament and governs the observed oscillations of sperm cells
around the centerline $\y=0$ of the filament. Heuristically, these
oscillations result from sperm cells slowly aligning their helix axis $\h$
parallel to $\Nabla_\perp c$ while approaching $\y=0$. At $\y=0$,
$\Nabla_\perp c$ changes its direction, yet sperm cells overshoot due to
their finite chemotactic turning speed $\bet<\infty$, before they
eventually make a `U-turn'. The second term
$\sign{\x} \gam \dt{\x}\dt{\y}$ in \eqref{dimless-ODE} originates from
the exponential decay of concentration along the centerline of the
filament and changes the amplitude of the oscillation. In particular,
for $\sign{\x\dt{\x}}<0$, i.e., sperm cells surfing towards the egg, the
oscillation is damped, whereas for $\sign{\x\dt{\x}}>0$, i.e., sperm cells
surfing away from the egg, it is amplified. This increase in amplitude
can cause sperm cells that are surfing away from the egg to eventually turn
around, redirecting them towards the egg. A linear stability analysis
of \eqref{dimless-ODE} around the case of a non-oscillating trajectory
$\leri{\y,\dt{\y}}=(0,0)$ yields the eigenvalues $\omega_{1,2}$ of the
Jacobian of the linearization,
\begin{linenomath}\begin{align}
\label{eq:frequency-damping-ratio}
  \omega_{1,2} = \zeta \pm \im \sqrt{1-\zeta^2} \ , \qquad \zeta =
  \sign{\x\dt{\x}}\frac{\gam}{2} \ ,
\end{align}\end{linenomath}
which define a harmonic oscillator with dimensionless damping ratio
$\zeta$ and dimensionless oscillation frequency $\sqrt{1-\zeta^2}$.
This analytic result agrees with full simulations of helical
chemotaxis in three-dimensional space, see
\FIG{damped-oscillation}.

Note that the predicted exponential decay of oscillation amplitude,
$\exp\leri{\zeta \t}=\exp\leri{\gam/2 \cdot t/\utau}$, is independent
of $x$ since $\gam/\utau$ is independent of $\sig^2(x)$.
Interestingly, both for \eqref{dimless-ODE} and full simulations, the
angle at which trajectories intersect the centerline $\y=0$ of the
concentration filament is essentially independent of the angle at
which they first entered the filament at $\ymax$, provided $\ymax$ is
sufficiently large: For smaller $\ymax$, i.e., outer and thus thinner
parts of the filament, trajectories will simply pass through the
filament, unable to execute a successful turn before they have left
the filament again. As the width of the filament decreases away from
the egg, this implies that filament surfing will be operative, at
most, up to a maximal distance from the egg (which depends on the
entry angle), characterized by $p_{\text{in}}$. If we account for
convection by shear flow $\vf = \shear y \, \e_x$,
\eqref{dimless-ODEx} changes to
$\dt{\x} \to \dt{\x} + \shear \utau (\y+y_0(\x)/L)$. Note that due to
$\sign{y_0(x)}=\sign{x}$, sperm cells that surf within the filament towards
the egg swim on average against the external flow.

\subsection{Minimal theory for sperm-egg-\successs}
\label{sec:effective-volume}

We provide an estimate for the \successs{} $\su$, building on the
effective equation of motion of the helix axis derived in
\SI{sec:chemotactic-navigation}. The \success{} $\suc$ is
obtained then from $\su$ using fertilization kinetics,
\eqref{fertilization-kinetics}. For $\su$, we decompose the search
problem for the egg into an outer search problem of finding the
concentration filament and an inner search problem of surfing along
the filament. We obtain (exploiting the symmetry between the two
branches of the filament for $x<0$ and $x>0$)
\begin{linenomath}\begin{equation}
\label{eq:encounter-prob}
  \su \approx 2 \Int_{0}^{\Rmax} \id{x} p_{\text{in}}(x,\tmax) \Big[A(x) \rho_{\text{egg}} + S(x) j_{\text{out}}
  t_{\text{out}}(x,\tmax) \Big] \ .
\end{equation}\end{linenomath}
Here, we introduce the following quantities:
\begin{itemize}
\item the cross-sectional area $A(x)$ at the position $x$ of the
  filament, which is defined by $c(\vec{r})\geq\cb$, i.e.,
  $A(x) = \iint_{-\infty}^{\infty} \id{y} \id{z}
  \Theta(c(x,y,z)-\cb)$,
  with the Heavyside-function $\Theta$ ($\Theta(c>0)=1$ and
  $\Theta(c\leq0)=0$),
\item the circumference $S(x)$ corresponding to the cross-section,
\item the average probability
$p_{\text{in}}(x,\tmax)$ that a trajectory entering the filament at
$x>0$ will surf along it and reach the egg within exposure time
$\tmax$,
\item the mean steady-state flux $j_{\text{out}}$ of trajectories
  arriving at the surface of the filament, and
\item the time limit $t_{\text{out}}$ for the outer search problem.
\end{itemize}
These quantities are explained in detail below. The first term in
\eqref{encounter-prob} accounts for sperm cells found inside the
concentration filament already at $t=0$, assuming a random uniform
distribution of initial positions. The second term in
\eqref{encounter-prob} accounts for trajectories, which first search
for the filament and, after encountering the filament, surf along it
towards the egg.

We compute the probability $p_{\text{in}}(x,\tmax)$ of successful
inner search numerically using the effective equation of motion for
the helix axis \eqref{dimless-ODE} as function of entry position $x$
and exposure time $\tmax$. Specifically, we average over simulations
of \eqref{dimless-ODE} with uniformly distributed initial entry points
and isotropic initial directions, i.e., entry angles. In order to
account for the ellipsoidal cross-section of the concentration
filament with $\sig_y=\sig\ratio$, $\sig_z=\sig$, we average results
for $\sig_y$ and $\sig_z$. From the successful trajectories, we also
obtain the mean travel time $t_{\text{in}}$ within the filament, which
represents a conditional mean first passage time. Accordingly, we set
the maximal time $t_{\text{out}}$ allowed for the outer search
$t_{\text{out}}(x,\tmax) = \tmax-t_{\text{in}}(x,\tmax)$ if
$p_{\text{in}}>0$ and $t_\text{out}=0$ else.

Note that the first term in \eqref{encounter-prob} can be written as
$V_{\text{eff}} \rho_{\text{egg}}$ with an effective volume
$V_{\text{eff}} = 2 \int_{0}^{\infty} \id{x} A(x)
p_{\text{in}}(x,\tmax)$
of the concentration filament, weighted by the probability
$p_{\text{in}}$ of successful chemotaxis to the egg. This contribution
is negligible compared to the second term for long exposure times
$\tmax$ and low egg densities $\rho_{\text{egg}}$.

The flux $j_{\text{out}}$ of trajectories arriving at the surface of
the concentration filament can be determined by a fit to $\su(\shear)$
from simulations at different \shearrate{}s $\shear$. Alternatively,
we can estimate $j_{\text{out}}$ by treating sperm cells outside of the
filament as ballistic swimmers with net swimming speed $\vs$ and
uniformly distributed random positions $\vec{r}$ and orientations $\h$
with probability distribution
$p_{\text{sperm}}(\vec{r},\h) = \leri{\frac{4}{3} \pi
  (\Rmax^3-\Regg^3)}^{-1} \leri{4 \pi}^{-1} \approx \rho_{\text{egg}}
\leri{4 \pi}^{-1}$.
Assuming that the filament is convex, each point on its surface is
reached at time $t$ from initial conditions on a surface of a
half-sphere with radius $\vs t$. The flux of trajectories with
direction $\h$ into the filament at $\vec{r}_0$ is
$j_{\text{out}}(\vec{r}_0,\h) = -\vec{n} \cdot \vs\h
p_{\text{sperm}}(\vec{r}_0,\h)$
for $\vec{n}\cdot\h<0$ and $j_{\text{out}}(\vec{r}_0,\h)=0$ else,
where $\vec{n}$ denotes the outer surface normal vector at
$\vec{r}_0$. For the constant density
$p_{\text{sperm}}(\vec{r}_0,\h)=p_{\text{sperm}}$ the total flux of
sperm cells into the filament is
$j_{\text{out}} = \int_0^{2\pi} \id{\phi} \int_0^{\pi/2} \id{\theta}
\sin \theta j_{\text{out}}(\vec{r}_0,\h(\phi,\theta)) =
p_{\text{sperm}} \pi \vs$,
where we use spherical coordinates $\phi, \theta$ with
$\vec{e}_z = \vec{n}$ to express $\h$. Note that an isotropic
distribution of orientations $\h$ is a simplification, since
co-rotation by flow alters this distribution, see
\SI{sec:jeff-eq-anal}.

Despite the simplifications made, \eqref{encounter-prob} can
quantitatively account for the \successs{} in full simulations, see
\FIG{suc}. In particular, we find that the numerical fit for
$j_{\text{out}}=0.063~\text{m}^{-2}\text{s}^{-1}$ is close to our
simple estimate for a ballistic swimmer
$j_{\text{out}} = \rho_{\text{egg}} \vs/4 =
0.04~\text{m}^{-2}\text{s}^{-1}$.
Of course, our simple theory has limitations: First, trajectories are
three-dimensional, not two-dimensional, and are characterized by
oscillations both in $y$- and $z$-direction. As a result, sperm
trajectories are super-helical, which reduces the effective speed
along the filament. Second, our theory does not account for the fact
that some sperm cells may miss the egg on the first attempt, and find it
only after reversing their motion in $x$-direction, which increases
the mean time $t_{\text{in}}$ to find the egg. Preliminary simulations
suggest that the difference between simulations and theory in
\FIG{suc} indeed originate from this effect. Finally,
co-rotation is neglected in the simple theory. However, this is
justified for $\shear \ll \utau^{-1}$, see \eqref{time-scale}, i.e.,
when rotation due to navigation is much faster than co-rotation due to
flow. Note that simulations with neither convection nor co-rotation
exhibit also an optimal \shearrate{} \shearopt{}, but at higher
\shearrate{} and different \successs{}. The reason is that convection
implies a flow opposing surfing towards the egg, which increases
$t_{\text{in}}$ compared to the case without convection. Thus, $\su$
increases for large $\shear$ when convection is not included,
resulting in a shift of $\shearopt$.

For the experiment of Zimmer and Riffell (data reproduced in
\FIG{fig5c} and \FIGSUPP{fig5c-coro}), we estimate a high
background concentration of chemotattractant
$c_{\text{bg}}\sim 4~\text{nM}$, see \SI{sec:constants}.
Adding a background concentration $c \to c + c_{\text{bg}}$ in
\eqref{filament-2D} leads to an effective, higher threshold
$\cb{}_{\text{,eff}} = \cb + c_{\text{bg}}$ in \eqref{dimless-ODE}.
Consequently, the volume of the filament with sufficiently high
concentration $c(\vec{r}) \geq \cb{}_{\text{,eff}}$ is situated only
in the vicinity of the egg. While our far-field theory of filament
surfing does not apply directly to this special near-field case, we
can make a simple estimate: We assume that sperm cells always swim directly
towards the egg within the concentration plume defined by
$c(\vec{r}) \geq \cb{}_{\text{,eff}}$ due to the close-to-spherical
shape of the plume. Thus, sperm cells entering the plume at $x_0=0$ approach
it with net radial speed $\vs$, as the external flow only convects the
sperm cells parallel to the egg surface, see \eqref{stokes-flow}. A second,
alternative calculation applies if sperm cells enter the plume at
$x_0\gg\Regg$: In this case, we can estimate the net speed towards the
egg by $\dt{x}=\shear y_0(x)-\vs$. This yields for the distance $x(t)$
from the egg,
$x(t) = \frac{\vs}{\shear b} + \leri{x_0 - \frac{\vs}{\shear b}}
\exp\leri{\shear b t}$
(using $y_0(x)\approx b x$, see \SI{sec:filament-anal}). We
use these two limit cases to compute $p_{\text{in}}$ and
$t_{\text{out}}$ for \eqref{encounter-prob} and obtain similar
fertilization probabilities $\suc(\shear)$ in both cases. For these
limit cases, $\suc(\shear)$ displays a similar decay as function of
$\shear$ as the simulation results without co-rotation, see
\FIG{fig5c}. In particular, the fitted flux
$j_{\text{out}} = 4.8\cdot10^3~\text{m}^{-2}\text{s}^{-1}$ is
consistent with the theoretical value
$j_{\text{out}} = \rho_{\text{egg}} \vs/4 =
7.5\cdot10^3~\text{m}^{-2}\text{s}^{-1}$.

\subsection{Analytic solution of Jeffery equation in shear flow}
\label{sec:jeff-eq-anal}

As shear flow is a fundamental paradigm for small-scale turbulence, we
present here the analytic solution to the Jeffery equation,
\eqref{jeff-eq}, for particles suspended in simple shear flow. The
application to helical swimmers is discussed. The results provide the
distribution of helix orientations $\h$ on the periodic boundary used
in the simulations, i.e., $p_{\text{sperm}}$ in \eqref{boundary}. In
particular, the results quantify the common notion that non-spherical
swimmers align their major axis parallel to the flow direction. In
fact, these swimmers rotate all the time, but with non-constant
rotation rate, causing these swimmers to spend more time aligned with
the flow axis. Consequently, the time-average of the orientation
vector is not zero, but aligned with the flow axis. Note that analytic
results for Poiseuille flow can be found in
\cite{ZoeSta2012,ZoeSta2013}.

For simple shear flow $\vf=\shear y \, \e_x$, the dynamics of the unit
vector $\e$ along the major axis of a prolate spheroid, i.e.,
$\dt{\e} = \vec{\Omega_{\text{f}}}\times \e$ with
$\vec{\Omega_{\text{f}}}$ given by \eqref{jeff-eq}, can be rewritten
in terms of spherical coordinates
$0\leq\theta\leq\pi, 0\leq\phi\leq2\pi$ of
$\e = (\sin\theta\cos\phi,\sin\theta\sin\phi,\cos\theta)$
\begin{linenomath}\begin{equation}
\label{eq:jeff-sphere}
\begin{aligned}
  \dt{\theta} &= \frac{\shear G}{4} \sin{2 \theta} \sin{2 \phi} \ ,
  \\
  \dt{\phi} &= \frac{\shear}{2} \lerii{G \cos{2 \phi} - 1} \ .
\end{aligned}
\end{equation}\end{linenomath}
The range $1\leq\geofactor<\infty$ of the aspect ratio $\geofactor$ (with
$\geofactor=0$ for a sphere and $\geofactor\to\infty$ for an infinitesimal
thin rod) implies $0\leq G < 1$ for the geometric factor $G$. The
dynamics of the polar angle $\phi(t)$ is independent of the azimuthal
angle $\theta(t)$. By integration, we find
\begin{linenomath}\begin{align}
\label{eq:phit}
    \phi(t) = \arctan\lerii{\frac{G-1}{\sqrt{1-G^2}} \tan\leri{\Psi(t)}}
\end{align}\end{linenomath}
with short-hand
\begin{linenomath}\begin{align}
\Psi(t) = \frac{\shear t}{2} \sqrt{1-G^2} + \arctan\lerii{\frac{\sqrt{1-G^2}}{G-1} \tan{\phi_0}}
\end{align}\end{linenomath}
and initial condition $\phi(0)=\phi_0$. Note that $\dt{\phi}\leq 0$,
i.e.,
$-\frac{\shear}{2} \leri{1+G} \leq \dt{\phi}
\leq-\frac{\shear}{2}\leri{1-G}$.
Hence, the polar angle $\phi(t)$ rotates clockwise with period
\begin{linenomath}\begin{align}
\label{eq:jeff-period}
T = \frac{4 \pi}{\shear \sqrt{1-G^2}}
\end{align}\end{linenomath}
with $T\geq 4\pi / \shear$. Substituting \eqref{phit} for $\phi(t)$
into \eqref{jeff-sphere}, we find
\begin{linenomath}\begin{align}
\label{eq:thet}
  \theta(t) = \arccot\lerii{\cot\leri{\theta_0}
  \sqrt{\frac{1+G \cos\lerii{2 \Psi(0)}}{1+G \cos\lerii{2 \Psi(t)}}}}
\end{align}\end{linenomath}
with initial condition $\theta(0)=\theta_0$.

We also compute the density $\rho_{\e}(\theta,\phi)$ of directions for
an ensemble of ballistic microswimmers obeying \eqref{jeff-sphere}.
The distribution of polar angles $\rho_{\phi}(\phi)$ is proportional
to $1/\leriii{\dt{\phi}}$
\begin{linenomath}\begin{align}
  \rho_{\phi}(\phi) = \frac{\sqrt{1-G^2}}{4 \pi \lerii{1 - G
  \leri{1-2\sin^2{\phi}}}} \ .
\end{align}\end{linenomath}
This density has two maxima, at $\phi_+=0$ and $\phi_+=\pi$, and two
minima at $\phi_-=\pm\pi/2$, resulting in a density range
$\rho_{\phi}(\phi_-)\leq \rho_{\phi} \leq \rho_{\phi}(\phi_+)$ with
$\rho_{\phi}(\phi_\pm)=\leri{4\pi}^{-1} \lerii{(1+G)/(1-G)}^{\pm
  1/2}$.

In order to derive the full density $\rho_{\e}(\theta,\phi)$, we use
an alternative scheme to solve the continuity equation, inspired by
the method of characteristics. Effectively, an ordinary differential
equation (ODE) and a system of ODEs are solved instead of one partial
differential equation (PDE). The dynamics of $\e$ correspond to a flow
$\vec{w}(\e)$ on the unit sphere. The continuity equation for a
density $\rho_\e(\e,t)$ in an arbitrary flow field $\vec{w}(\e,t)$
reads \begin{linenomath}\begin{align} \dx{\rho_\e(\e,t)}{t} = - \Nabla \cdot
  \lerii{\rho_\e(\e,t) \vec{w}(\e,t)} \ .
\end{align}\end{linenomath}
Instead of solving directly for the density $\rho_\e(\e,t)$ in the
laboratory frame, we can first solve for the density
$\breve{\rho}_\e\leri{t; \e_0}$ in a co-moving frame
\begin{linenomath}\begin{align}
  \rho_\e(\e,t) = \breve{\rho}_\e\leri{t; \vec{\breve{\e}}(-t,\e)}
  \ ,
\end{align}\end{linenomath}
where $\breve{\e}(t;\e_0)$ is the trajectory starting at
$\breve{\e}(0;\e_0)=\e_0$ and following the flow
$\dt{{\breve{\e}}} = \vec{w}(\vec{\breve{\e}},t)$. We obtain
$\breve{\rho}_\e\leri{t; \e_0}$ from the rewritten continuity equation
\begin{linenomath}\begin{align}
  \dt{\breve{\rho}}_{\e}\leri{t;\e_0} = -
  \breve{\rho}_{\e}\leri{t;\e_0} \Nabla \cdot
  \vec{w}\leri{\breve{\e}(t;\e_0),t} \ .
\end{align}\end{linenomath}
Applying this scheme to \eqref{jeff-sphere} with flow
$\vec{w}(\theta, \phi) = \dt{\theta} \e_\theta + \sin \theta \dt{\phi}
\e_\phi$
on the unit sphere and using the solutions $\theta(t), \phi(t)$ from
Eqs.~(\ref{eq:phit},\ref{eq:thet}) yields
\begin{linenomath}\begin{equation}
\begin{split}
  \breve{\rho}_{\e}(t; \theta_0, \phi_0) = C_0(\theta_0, \phi_0) &
  \left[ \cot^2\leri{
      \theta_0} \leri{1+G \cos\lerii{2 \Psi(0)}} \right. \\
  & \left. \vphantom{\cot^2} + \leri{1+G \cos\lerii{2 \Psi(t)}}
  \right]^{\frac{3}{2}} \ ,
\end{split}
\end{equation}\end{linenomath}
where the pre-factor $C_0(\theta_0, \phi_0)$ is defined by the initial
conditions. For our simulations, we use an initially uniform
distribution such that
$\breve{\rho}_{\e}(0,\theta_0,\phi_0) = \leri{4\pi}^{-1}$. Switching
notation to $\breve{\theta}(t, \theta_0, \phi_0) = \theta(t)$ and
$\breve{\phi}(t,\phi_0)=\phi(t)$, the density $\rho_{\e}$ follows
\begin{linenomath}\begin{align}
  \rho_{\e}(\theta, \phi, t) = \breve{\rho}_{\e}\leri{t;
  \breve{\theta}(-t,\theta,\phi), \breve{\phi}(-t,\phi)} \ .
\end{align}\end{linenomath}
While $\rho_{\e}$ is periodic in time with period $T$ by
\eqref{jeff-period}, we can compute a time-average over one period,
starting with a uniform distribution of directions $\e$ at $t=0$. The
time-averaged density displays a maximum at the axis of flow
$\e = \pm \e_x$ and a minimum at the shear axis $\e = \pm \e_y$. These
extrema vanish for a sphere ($G=0$) and become more pronounced with
increasing $G$.

While the above results are derived for the case of a suspended
particle, numerical simulations show that they also approximately
apply to the centerline $\vec{r}_h(t)$ of a helical swimmer with helix
axis $\h$ (without chemotaxis) if we use an effective aspect ratio
$\geofactor_{\text{eff}}$. Specifically, the dynamics of the helix axis
$\h$ resembles the above solutions with a smaller aspect ratio
$1\leq \geofactor_{\text{eff}} \leq \geofactor$. This approximation is valid
for small times $t$ and at small $\shear$, i.e., as long as the helix
period is much smaller than the period $T(\geofactor)$. For instance, we
fit $\geofactor_{\text{eff}}=1.3\pm 0.1$ ($G_{\text{eff}}=0.26$) for the
sea urchin helix parameters and $\geofactor=5$ ($G=0.92$). This effective
parameter is a result of averaging the instantaneous co-rotation for
the swimming direction $\e_1$ with parameter $G$ over one period of
helical swimming. Generally, $\geofactor_{\text{eff}}$ depends on the
angle between $\h$ and $\e_1$. For larger $\shear$, complicated
behavior of $\h$ is observed with limit cycles and stable fixed
points, which is consistent with recent results for Jeffery equation
in perturbed shear flow~\cite{SloAlcSecStoFer2020}. We use the value
$\geofactor_{\text{eff}}$ in all simulations to determine the periodic
boundary conditions at the boundary of the simulation domain.

\subsection{Choice of parameters}
\label{sec:constants}

Parameters used throughout the three simulation scenarios
(\emph{Arabacia punctuala} for \FIG{filament-surfing}B,
\FIG{suc}, \FIG{damped-oscillation}, \emph{Strongylocentrotus
  purpuratus} from
\cite{MeaDen1995,Mea1996,DenNelMea2002,Gay2008} for
\FIG{fig5}, \FIGSUPP{fig5-coro} and \FIGSUPP{fig4},
\emph{Haliotis rufescens} from \cite{RifZim2007, ZimRif2011} for
\FIG{fig5c} and \FIGSUPP{fig5c-coro}) are listed in
Table~\ref{tab:constants} and discussed in the following.

Mean path curvature $\curv_0=0.065~\mu\text{m}^{-1}$ and mean path
torsion $\tor_0=0.067~\mu\text{m}^{-1}$ of the helical paths are set
according to three-dimensional tracking of \emph{A. punctuala}
sperm cells~\cite{JikAlvFriWilPasColPicRenBreKau2015}. Three-dimensional
tracking for \emph{S. purpuratus} give similar
values~\cite{CorTabWooGueDar2008}, though with larger error intervals.
Moreover, the sperm morphology for \emph{A.
  punctuala}~\cite{JikAlvFriWilPasColPicRenBreKau2015,KroMaeLanBaiFri2018},
\emph{S. purpuratus}~\cite{AlvDaiFriKasGrePasKau2012}, and \emph{H.
  rufescens}~\cite{HorBenBis2018} is similar, which justifies the use
of the same helix parameters for all three species. Likewise, the
effective aspect ratio $\geofactor=5$ between major and minor axis of a
sperm cell, i.e., length of flagellum divided by typical beat
amplitude, suggested for \emph{H. rufescens}~\cite{RifZim2007} is
employed for all three species in the Jeffery equation
\eqref{jeff-eq}. We observe that simulation results are largely
independent of the precise value of $\geofactor$. The signaling time-scale
$\mu = 1/\leri{\vo \sqrt{\curv_0^2 +\tor_0^2}}$ is chosen to ensure
the optimal phase-lag between concentration input $c(\vec{r}(t))$ and
motor response $a(t)$~\cite{FriJue2007,Fri2008}, see
\eqref{eq-output}, consistent with experimental
observations~\cite{JikAlvFriWilPasColPicRenBreKau2015}. For all three
species, the gain factor is set as $\psperm=5$, corresponding to the
mean of the values used in \cite{KroMaeLanBaiFri2018}. This value
reproduces typical bending rates of helical swimming paths as observed
in experiments~\cite{JikAlvFriWilPasColPicRenBreKau2015}. The
threshold of sensory adaption $\cb=10~\text{pM}$ is chosen as
suggested in \cite{KasAlvSeiGreJaeBeyKraKau2012}. At the
concentration $c_b$, about $20$ chemoattractant molecules would
diffuse to a sperm cell during one helical turn. Note that sea urchin
sperm cells respond to single chemoattractant
molecules~\cite{Picetal2014}; the change in intra-cellular calcium
concentration caused by the binding of chemoattractant molecules as
function of stimulus strength becomes sub-linear already for
chemoattractant concentrations on the order of
$c_b$~\cite{KasAlvSeiGreJaeBeyKraKau2012}. For \emph{A. punctuala},
other parameters were also tested, i.e., $\psperm=2$ and
$\cb=1~\text{pM}$, which yielded qualitatively similar simulation
results and again agreement of theory and simulations. Note that the
experimental protocol used in \cite{ZimRif2011} for \emph{H.
  rufescens} results in a substantial background concentration of
chemoattractant, which we estimate as $c_{\text{bg}}\sim 4~\text{nM}$
(experiments are conducted $10-30~\text{min}$ after spawning at a high
density of eggs $\rho_{\text{egg}} = 10^3~\text{ml}^{-1}$ with the
known release rate $\relrate = 0.18~\text{fmol min}^{-1}$ of
chemoattractant~\cite{RifKruZim2004}). According to our theory, such a
background concentration causes effectively a higher sensitivity
threshold $\cb{}_{\text{,eff}} = \cb + c_{\text{bg}}$ (see
\SI{sec:effective-volume}), which may the be reason for the
higher behavioral threshold $300~\text{pM}$ observed in
\cite{ZimRif2011}. In the case of \emph{S. purpuratus}, we
estimate an even higher background concentration,
$c_{\text{bg}}\sim 500-4000~\text{nM}$, which renders chemotaxis
ineffective. For this estimate, we use that experiments were conducted
$1-8~\text{h}$ after spawning at a high egg density
$\rho_{\text{egg}} = 1.5\cdot
10^4~\text{ml}^{-1}$~\cite{MeaDen1995,Mea1996}
and assume a release rate $\relrate = 0.46~\text{fmol min}^{-1}$ of
chemoattractant as for \emph{A.
  punctuala}~\cite{KasAlvSeiGreJaeBeyKraKau2012}.

For the swimming speed $\vo$ of sperm cells along helical paths for both sea
urchin species, we use the measured value $\vo=200~\mu\text{m s}^{-1}$
from \cite{JikAlvFriWilPasColPicRenBreKau2015}. Note that some
experiments effectively measure the net swimming speed along the helix
axis $\vs = \vo\tor_0 /\sqrt{\curv_0^2+\tor_0^2}$, which is smaller
than $\vo$. For \emph{H. rufescens}, we use the speed $\vs$ measured
during the same experiment~\cite{ZimRif2011}. Note that this
experiment also indicated chemokinesis, i.e., higher swimming speeds
at elevated chemoattractant concentration, an effect which we neglect
here for simplicity.

For \emph{A. punctuala}, we use the diffusion coefficient
$\D=239~\mu\text{m}^2 \text{s}^{-1}$ and release rate
$\relrate=0.46~\text{fmol min}^{-1}$ of
chemoattractant~\cite{KasAlvSeiGreJaeBeyKraKau2012}. For this
simulation, we assume a low egg density
$\rho_{\text{egg}}=10^{-3}~\text{ml}^{-1}$, which yields the radius
$\Rmax=6\cdot10^4~\mu\text{m}$ of the outer boundary centered around
the egg according to
$\rho_{\text{egg}} = \leri{4 \pi \Rmax^3 / 3}^{-1}$. For this
reference case, the filament is completely included inside the
simulation domain for all considered \shearrate{}s $\shear$. The
exposure time $\tmax=360~\text{s}$ is chosen comparable to the
experiment in \cite{MeaDen1995}, where $\tmax=120~\text{s}$.
While for this work the exposure time $\tmax$ is set by the protocol
of the considered experiment, in a generic turbulent flow $\tmax$
corresponds to the time-scale of flow changes, i.e. scale with the
Kolmogorov time $\tmax \sim \tkol$. For comparison with the
experiments with \emph{S. purpuratus} and \emph{H. rufescens}, the
radius $\Rmax$ is computed directly from the stated egg density
$\rho_{\text{egg}} = \leri{4 \pi \Rmax^3/3}^{-1}$. From the
$5~\text{vol}\%$ solution with $\Regg=40-55~\mu$m~(\cite{Mea1996}, pg.
161), we infer a range
$\rho_{\text{egg}} = 0.9-3.4 \cdot 10^4~\text{ml}^{-1}$ for the
experiments with \emph{S. purpuratus}. This estimate already takes
into account that, according to the experimental protocol, the above
egg solution is mixed $9:1$ with sperm solution~\cite{MeaDen1995,
  Mea1996}. Likewise, from the range of sperm densities
$\rho_{\text{sperm}} = 1.9-3.1 \cdot 10^6~\text{ml}^{-1}$ in
Fig.~4 of \cite{MeaDen1995} and the estimate
$\rho_{\text{sperm}}=4\cdot 10^6~\text{ml}^{-1}$ from pg.~59 of \cite{Mea1996}, both before $9:1$-dilution, we infer a final
concentration $\rho_{\text{sperm}}=3.9 \cdot 10^5~\text{ml}^{-1}$. We
use the kinematic viscosity $\nu = 10^{-6}~\text{m}^2\text{s}^{-1}$ of
sea water at room temperature.

\subsection{Numerical Simulation}
\label{sec:numerical-simulation}

The equations of motion are integrated using an Euler scheme with
fixed time step $\id{t}$. For all time integrations, a time step
$\id{t} = 10^{-3}~\text{s}$ is used. Integration with smaller
$\id{t}=10^{-4}~\text{s}$ for some test cases gave consistent results.
The number $N_{\text{sperm}}$ of sperm cells simulated in each case is
$10^5$, except for \emph{S. purpuratus}, where $N_{\text{sperm}}=10^4$
is used.

The concentration field is computed from Lagrangian particle tracking
with Euler-Maruyama method for the Fokker-Planck equation
\begin{linenomath}\begin{align}
  \partial_{t} c = - \Nabla \cdot \vf \ c
  + \D \bigtriangleup c
\end{align}\end{linenomath}
with $\vf$ from \eqref{stokes-flow}.
Test particles were released at random points of the surface of sphere of radius $\Regg$ located at the origin.
In total, we used $4\cdot 10^6$
test particles, which corresponds effectively to $1.6 \cdot 10^7$
particles by exploiting symmetries of the flow field. Concentrations
are evaluated on a cubic $50 \times 50 \times 50$ grid, spanning in
each dimension from $-\Rmax$ to $\Rmax$, and then interpolated by a
spline interpolation of order 3. This grid is sufficiently fine to
resolve the details of the concentration filaments. The rapid
convergence to a near-steady state allows to use a static
concentration field corresponding to exposure time $\tmax$ for each
simulation. We checked for test cases that full simulations with
time-varying concentration field do not yield different results.

The implementation of an unsteady shear flow for a \shearrate{}
$\shear$ used as illustration in \FIG{filament-surfing}A is
inspired by \cite{BelCri2015}: We use the flow field
$\vf(\vec{r},t) = \shear'(r,t) \ \lerii{\vec{r} \cdot \e'_y(t)} \
\e'_x(t)$,
where the shear axis $\e'_y(t)$ and the flow axis $\e'_x(t)$ are
subject to a three-dimensional random walk on the unit sphere with
rotational diffusion coefficient $D_{\text{rot}} = \pi \shear$. The
\shearrate{} profile is given by
$\shear'(r,t) = \sqrt{2} \ \shear \sin\leri{2 \pi t/T_\shear} \ h(r)$.
The \shearrate{} $\shear'(r,t)$ decays as $h(r)$ with distance $r$
away from the center. This decay $h(r)$ mimics the decay of velocity
from the center of a vortex. We use the decay of an Lamb-Oseen vortex
$h(r)=\leri{\frac{r_{\text{core}}}{r}}^2
\leri{1-\exp\lerii{-\leri{\frac{r}{r_{\text{core}}}}^2}}$,
employing the Burger radius $r_{\text{B}}$ of a Burger vortex as core
radius $r_{\text{core}}= r_{\text{B}}$, where
$r_{\text{B}} = K \nkol \approx K \sqrt{\frac{\nu}{\shear}}$ with
$K=7.1$~\cite{HatKam1997,JumTroBosKar2009,WebYou2015}. The
\shearrate{} $\shear'(r,t)$ oscillates in time with root-mean-square
amplitude
$\sqrt{\frac{1}{T_\shear}\Int_0^{T_\shear} \id{t} \shear'(r,t)^2} =
h(r) \shear$
and period $T_\shear=\frac{r_B^2}{2 \nu}$, corresponding to the time
scale of decay of a Burger vortex.

\begin{table*}

\caption{\label{tab:constants}
  \textbf{List of parameters used or obtained for the three scenarios.} See text
  for discussion and further parameters.}
\setlength{\hsize}{15cm}
\begin{tabular*}{\hsize}{p{0.25\hsize} P{0.05\hsize} p{0.12\hsize}| P{0.15\hsize} P{0.2\hsize}
  P{0.15\hsize}}
  Parameter & & & Sea urchin~\cite{JikAlvFriWilPasColPicRenBreKau2015}
                  \newline (\emph{A. punctuala}) \newline
                  \FIG{filament-surfing}, \FIG{suc}, \FIG{damped-oscillation}  & Sea urchin \cite{MeaDen1995,Mea1996,DenNelMea2002,Gay2008}
                                                                                                   \newline
                                                                                                   (\emph{S.
                                                                                                   purpuratus})
                                                                                                   \newline
                                                                                                   \FIG{fig5},
                                                                                                   \FIGSUPP{fig5-coro},
                                                                                                   \FIGSUPP{fig4}

  & Red abalone~\cite{RifZim2007,
    ZimRif2011}
    \newline
    (\emph{H.
    rufescens}) \newline \FIG{fig5c}, \FIGSUPP{fig5c-coro} \\[1mm]
  \cline{1-6}
  path curvature & $\curv_0$ & $\lerii{\mu\text{m}^{-1}}$  &
                                                             $0.065$ &
  &
  \\
  path torsion & $\tor_0$ & $\lerii{\mu\text{m}^{-1}}$  &
                                                          $0.067$ & &
  \\
  helix radius & $r_0$ & $\lerii{\mu\text{m}}$ &
                                                 \multicolumn{3}{c}{$\curv_0/\leri{\curv_0^2
                                                 +\tor_0^2}\approx 7$} \\
  gain factor & $\psperm$ & & \multicolumn{3}{c}{$5$} \\
  threshold of sensory adaption & $\cb$ & $\lerii{\text{pM}}$ &
                                                                \multicolumn{3}{c}{$10$} \\
  signaling time-scale & $\mu$ & $\lerii{\text{s}}$ &
                                                      \multicolumn{3}{c}{$\leri{\vo
                                                      \sqrt{\curv_0^2
                                                      +\tor_0^2}}^{-1}$} \\
  sperm aspect ratio & $\geofactor$ & &
                                    \multicolumn{3}{c}{$5$} \\
  \cline{4-6}
  swimming speed & $\vo$ & $\lerii{\mu\text{m} \text{ s}^{-1}}$ &
                                                                  $200$ & \multicolumn{1}{c|}{}
  &
    $42$
  \\
  net speed along helix axis & $\vs$ & $\lerii{\mu\text{m}
                                       \text{ s}^{-1}}$ &
                                                          $145$ & \multicolumn{1}{c|}{}
  & $30$ \\
  chemoattractant release rate & $\relrate$ & $\lerii{\text{fmol min}^{-1}}$ &
                                                                               $0.46$
                                                                                                 &
                                                                                                   \multicolumn{1}{c|}{}& $0.18$  \\
  diffusion coefficient & $\D$ & $\lerii{\mu\text{m}^2 \text{s}^{-1}}$ &
                                                                         $239$
                                                                                                 &
                                                                                                   \multicolumn{1}{c|}{}
  & $660$ \\
  \cline{4-6}
  egg radius & $\Regg$ & $\lerii{\mu\text{m}}$ & $100$ & \multicolumn{1}{|c|}{$50$} & $108$ \\
  egg density & $\rho_{\text{egg}}$ & $\lerii{\text{ml}^{-1}}$ & $10^{-3}$ &
                                                                             \multicolumn{1}{|c|}{$1.5
                                                                             \cdot
                                                                             10^{4}$}
  & $10^{3}$
  \\
  boundary radius & $\Rmax$ & $\lerii{\mu\text{m}}$ & $6\cdot10^4$ &
                                                                     \multicolumn{1}{|c|}{$240$}
  &
    $620$ \\
  sperm density & $\rho_{\text{sperm}}$ & $\lerii{\text{ml}^{-1}}$ &
                                                                     --
                                                                                                 &
                                                                                                   \multicolumn{1}{|c|}{$3.9\cdot10^5$}
  & $10^4$
  \\
  exposure time & $\tmax$ & $\lerii{\text{s}}$ & $360$ & \multicolumn{1}{|c|}{$120$}
  &
    $15$ \\
  background concentration & $c_{\text{bg}}$ &
                                               $\lerii{\text{nM}}$
                & -- & \multicolumn{1}{|c|}{$500-4000$} & $4$ \\
  fertilizability (fit) & $\sucfert$ & & -- & \multicolumn{1}{|c|}{$10
                                              \%$} & $60 \%$ \\
\end{tabular*}
\end{table*}

\subsection{Parameter study}
\label{sec:parameter-variation}

\begin{figure}[h!]
  \hspace*{-4cm}\includegraphics{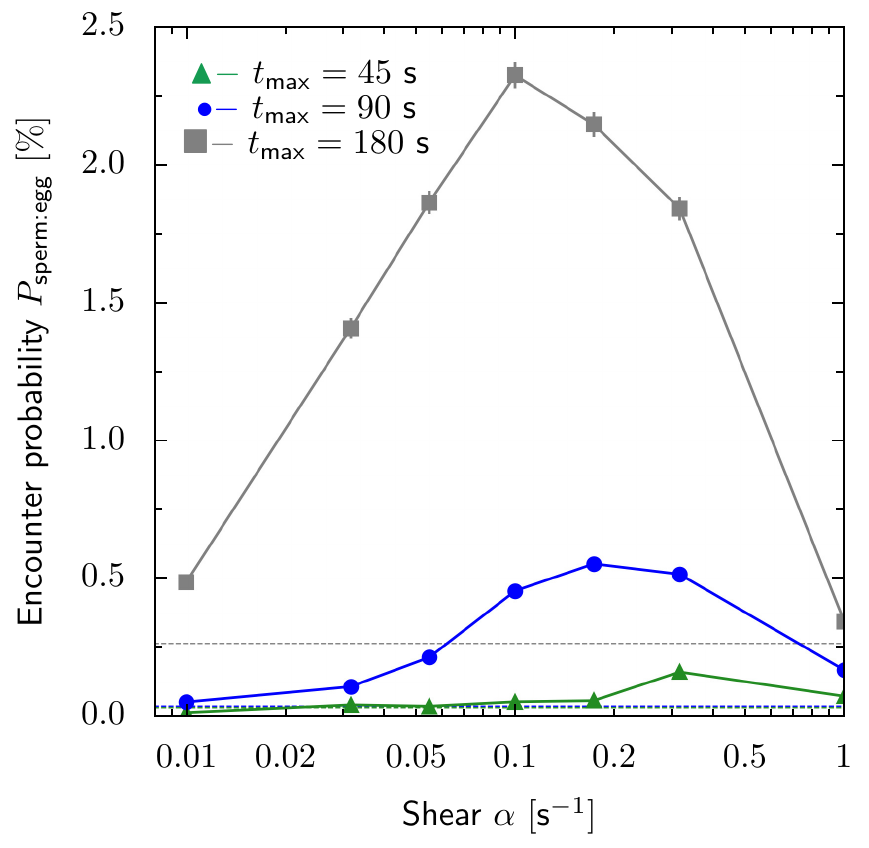}
\caption{\label{fig:tmax-variation} \textbf{Flow-dependent sperm-egg
    encounter probability for different exposure times $\tmax$.}
  Encounter probabilities $\su(\shear)$ as function of external
  \shearrate{} $\shear$ for three values of sperm-egg exposure time
  $\tmax$ obtained from simulations with co-rotation. (symbols
  according to legend, mean $\pm$ SD; flow-less results
  $\su(\shear = 0~\text{s}^{-1})$ displayed by dashed horizontal lines
  in respective color). Parameters taken for \emph{A. punctuala}, see
  Table~\ref{tab:constants}, except boundary radius $\Rmax=15~$mm for
  numerical efficiency.}
\end{figure}

\begin{figure}[h!]
\hspace*{-1.5cm}\includegraphics{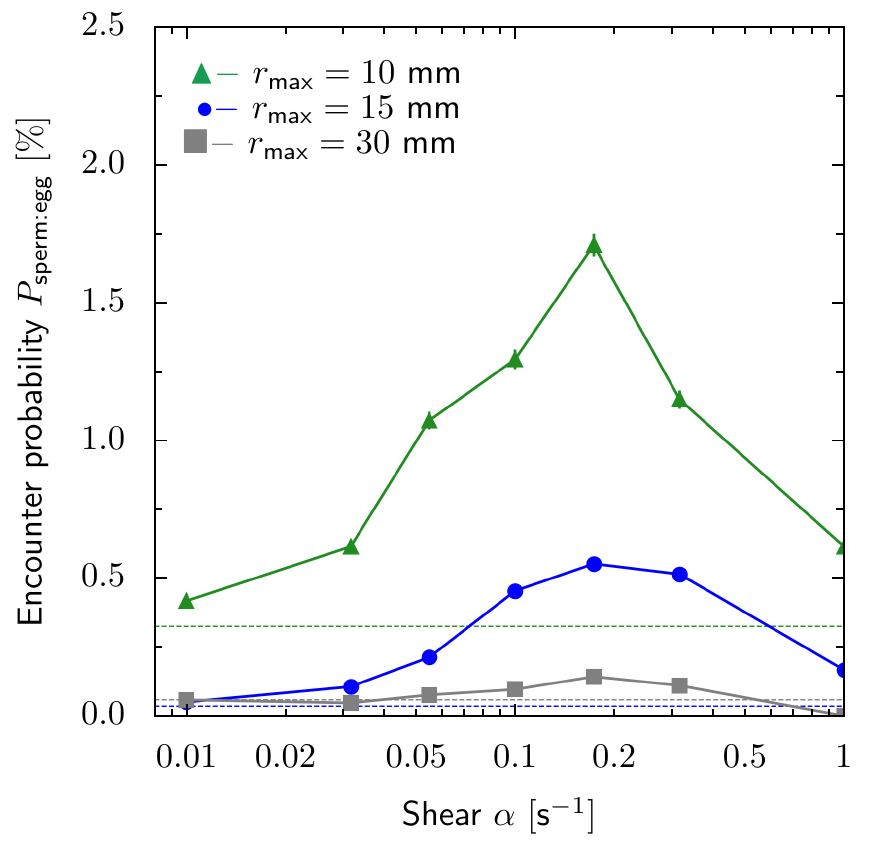}
\caption{\label{fig:Rmax-variation} \textbf{Flow-dependent sperm-egg
    encounter probability for different egg densities
    $\rho_{\text{egg}}$.} Analogous to \FIG{tmax-variation},
  yet for three different values of boundary radius $\Rmax$,
  corresponding to three different egg densities $\rho_{\text{egg}}$
  according to $\rho_{\text{egg}} = \leri{4 \pi \Rmax^3/3}^{-1}$. (For
  all three curves $\tmax=90~$s, thus blue curve identical to blue
  curve in \FIG{tmax-variation}.)}
\end{figure}

\begin{figure}[h!]
\hspace*{-4cm}\includegraphics{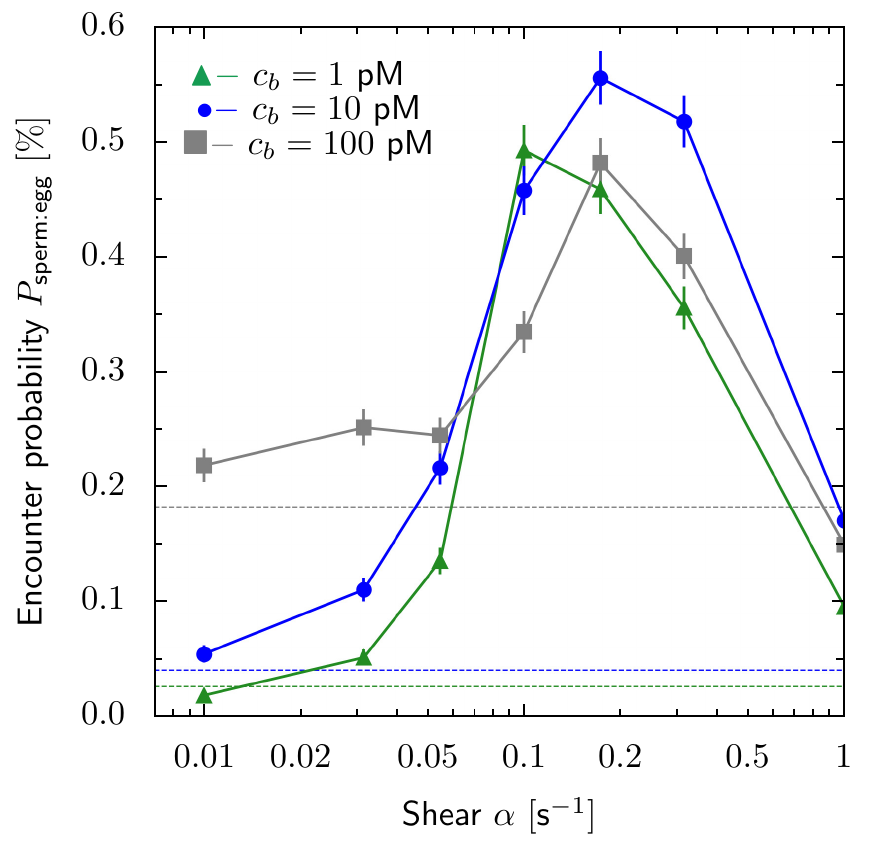}
\caption{\label{fig:sb-variation} \textbf{Flow-dependent sperm-egg
    encounter probability for different thresholds of sensory adaption
    $\cb$.} Analogous to \FIG{tmax-variation}, yet for three
  different values of threshold $\cb$. (For all three curves $\tmax=90~$s and $\Rmax=15$~mm,
  thus blue curve identical to blue curve in
  \FIG{tmax-variation}.)}
\end{figure}

\begin{figure}[h!]
\hspace*{-1.5cm}\includegraphics{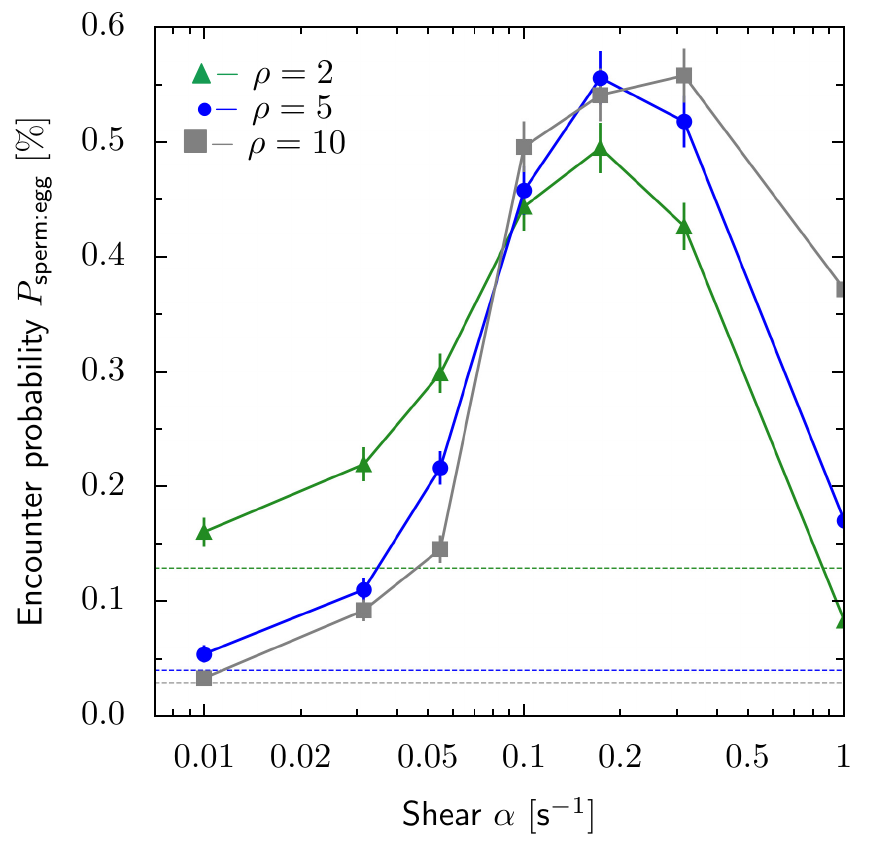}
\caption{\label{fig:rho-variation} \textbf{Flow-dependent sperm-egg
    encounter probability for different gain factors $\psperm$.}
  Analogous to \FIG{tmax-variation}, yet for three different
  values of gain factor $\psperm$. (For all three curves $\tmax=90~$s
  and $\Rmax=15$~mm, thus blue curve identical to blue curve in
  \FIG{tmax-variation}.)}
\end{figure}

In order to demonstrate the sensitivity of the quantitative results,
shown in \FIG{suc} of the main text, on
the parameters, we computed the \successs{} $\su(\shear)$ for a range
of exposure times $\tmax$ , egg densities $\rho_{\text{egg}}$
expressed in terms of boundary radii $\Rmax$, threshold of sensory
adaption $\cb$, and gain factors $\psperm$ as shown in
\FIG{tmax-variation}, \FIG{Rmax-variation},
\FIG{sb-variation}, and \FIG{rho-variation}, respectively. In
all cases, there is a pronounced optimum present at some intermediate
\shearrate{} $\shearopt$, where the position of the optimum $\shearopt$ is only slightly affected
by the parameter variations. The parameters mostly affect the height of
the optimum, in particular $\tmax$ and $\Rmax$, and its ratio to the
flow-less $\shear=0$ case, see Table~\ref{tab:maxratio}. This suggests
that the existence of an optimum is quite insensitive to parameter
variations, i.e., regardless of how the parameters are adapted,
fertilization is optimal at an intermediate shear flow for a broad
physiological range of parameter values.

For each parameter study all parameters but one are kept constant on
the values reported for \emph{A. punctuala} in
Table~\ref{tab:constants}. The only exception are $\tmax$ and $\Rmax$
whose base values are lowered to $\tmax=90$~s and $\Rmax=15$~mm for
numerical efficiency, i.e., the blue dots in
\FIG{tmax-variation}, \FIG{Rmax-variation},
\FIG{rho-variation}, and \FIG{sb-variation} correspond always
to the same parameters. The increase of the exposure time $\tmax$ in
\FIG{tmax-variation} from $45$ to $90$~s causes only a slight
decrease of the optimal \shearrate{} $\shearopt$ from $0.3$ to
$0.1~\text{s}^{-1}$, but increases the absolute \successs{}
$\su(\shearopt)$ by an order of magnitude. Such an increase of
$\su(\shearopt)$ is also observed for the increase of the egg density,
i.e., the decrease of the boundary radius from $30$ to $10$~mm in
\FIG{Rmax-variation}. These increases are in accordance with
the simple argument that longer search time or smaller search volume
increases the chances of finding the egg. The advantage of the optimum
to the flow-less case $\su(\shearopt)/\su(\shear=0)$
varies for both parameters between a factor 2 and 14, see
Table~\ref{tab:maxratio}. In contrast, the variation of $\cb$ and
$\psperm$ hardly affects the optimum in terms of $\shearopt$ and
$\su(\shearopt)$ but rather alters the probability in the absence of flow:
Increasing $\cb$ or decreasing $\psperm$ increases
$\su(\shear = 0~\text{s}^{-1})$. This is probably an effect of
signal-noise, originating from the computed concentration field which, due to
the very nature of Lagrangian particle tracking,
can exhibit low signal-to-noise ratio at low concentrations, i.e. at
the surface of the concentration plume. (Note that our model does not
explicitely account for sensing noise~\cite{BerPur1977}.)
This noise results in an effective reflection of incoming sperm
trajectories at the surface of the plume for increasing sensitivity of
the concentration measurement, expressed by $\cb$, or increasing
reaction to signal stimulus, expressed by $\psperm$, see also
discussion in \cite{KroMaeLanBaiFri2018,KrolaFri2020}. The effect is expected
to be much smaller for concentration filaments at
$\shear>0~\text{s}^{-1}$ as the concentration gradient towards the
center of the filament is higher and thus the signal-to-noise ratio
generally higher as for a concentration plume solely established by diffusion in the
flow-less case.

\vspace*{3cm}

\clearpage

\vspace*{10cm}

\begin{table*}
\centering
\caption{\label{tab:maxratio}
\textbf{Relative amplitude of optimum in sperm-egg encounter probability.}
Ratios of \successs{}
  $\su(\shear=\shearopt)$ at optimal \shearrate{} $\shearopt$
  normalized by \successs{} $\su(\shear=0)$ in the absence of flow
  for parameter study displayed in
  \FIG{tmax-variation}, \FIG{Rmax-variation},
  \FIG{sb-variation}, and \FIG{rho-variation}.}
\hspace{-2cm}
\begin{tabular*}{0.85\hsize}{l|P{0.1\hsize}P{0.1\hsize}P{0.1\hsize}}
  Parameter &
              \multicolumn{3}{c}{$\su(\shearopt)/\su(\shear=0)$} \\
  \hline
  \hline
  Sperm-egg exposure time: & & & \\
  $\tmax=$ 45 s, 90 s, 180 s & 4.6 & 13.6 & 8.7 \\[1mm] 
  \hline
  Boundary radius setting egg density: & & & \\
  $\Rmax=$ 10 mm, 15 mm, 30 mm & 5.2 & 13.6 & 2.3 \\[1mm] 
  \hline
  Threshold of sensory adaptation: & & & \\
  $\cb=$ 1~pM, 10~pM, 100~pM & 18.3 & 13.6 & 2.6 \\[1mm] 
  \hline
  Chemotactic gain factor: & & & \\
  $\psperm=$ 2, 5, 10 & 3.8 & 13.6 & 18.6 \\ 
\end{tabular*}
\end{table*}

\end{document}